# Performance analysis of a non-beacon enabled IEEE 802.15.4 network with retransmission and ACK mode

Yang Chen*[,1,3], Zhong-yi Wang[1,2,3], Lan Huang* [1,2,3]

[1]*College of Information and Electrical Engineering, China Agricultural University, Beijing 100083,China, e-mail: yancychy@163.com*

[2]*Modern Precision Agriculture System Integration Research Key Laboratory of Ministry of Education, Beijing 100083, China*

[3]*Key Laboratory of Agricultural Information Acquisition Technology (Beijing), Ministry of Agriculture, Beijing 100083, China, e-mail: biomed_hl@263.net*

**Abstract:** The aim of our work is to evaluate the performance of non-beacon IEEE 802.15.4 networks with acknowledgement (ACK) mode and retransmission limits in a finer time unit. Moreover, we predict network performance parameters using backpropagation artificial neural networks (BP-ANNs) with respect to a given quality of service for real-world applications. Therefore, our proposed methods can assist the deployment of a star network with its performance specified in a more practical way.

First, the discrete time Markov chain model and M/M/1/k queue are used to describe the full unslotted carrier sense multiple access with collision avoidance (CSMA/CA) algorithm in a non-beacon network. Considering MAC buffer size, unsaturated traffic, and saturated traffic, we build three analytical models to derive eight important performance metrics, e.g., throughput, delay, and reliability. In addition, extensive simulation results show the accuracy of the analytical models in throughput and reliability.

Finally, we use the analytical data to train the BP-ANNs models to predict the key parameters such as node number and delay. All results from the simulation data used to test the BP-ANNs show the accuracy these models. Thus, these methods and results can be used to deploy star networks in application environments.

**Acknowledgements**

This research was supported by the National Key Scientific Instrument and Equipment Developme

nt Projects (2011YQ080052), and the Specialized Research Fund for the Doctoral Program of Higher Education (20130008110035). The authors would like to thank the Key Laboratory of Agricultural Information Acquisition Technology of the Chinese Ministry of Agriculture for their support. Thanks are also due to Professor Rui-zhi Sun for providing the additional wireless sensor nodes in this research.

# 1. Introduction

IEEE 802.15.4 is an important standard in wireless networks. It defines the standard of data communication in wireless personal area networks with low data rate, low power, low complexity, and short-range radio frequency transmissions. It contains medium access control (MAC) and physical layer (PHY) communication protocols, and constitutes the fundamental protocol in ZigBee [1] and 6LowPan [2], which have been applied in smart homes, environment monitoring, healthcare, agriculture monitoring, and industry monitoring.

The key performance metrics of wireless sensor networks are throughput, reliability, and delay. Throughput reflects the utilization of the wireless channel as well as its congestion level. In theory, the maximum effective data rate of a single node in a wireless sensor network can reach 127 kbps [3]. Burchfield et al. reported that the maximum throughput in theory is 120 kbps, in simulation is 120 kbps, and in real hardware is 110 kbps [4]. The maximum throughput of 163 kbps or an efficiency of 64.9% can be achieved in the 2.4 GHz band [5]. Lee found that the presence of packet overheads, random backoffs, and acknowledgement (ACK) messages, as well as a large setting for inter-frame spacing are the reasons that the data rate is below the theoretical value of 250 kbps [6].

The transmission delay and packet delivery ratio are important metrics in industry control and healthcare applications. Low latency and high delivery ratios can ensure data transmission in real-time or near real-time. Zhu et al. analyzed the packet delay performance of an unsaturated non-ACK IEEE 802.15.4 beacon network with a finite buffer [7]. The results show that packet delay increases with buffer size. In addition, IEEE 802.15.4 MAC is not suitable for time-critical industry applications, as it is not able to guarantee an acceptable reliability level that is subject to the required timeliness [8]. The increase in the delivery ratio is in accordance with a significant higher latency. The authors of [8] therefore proposed an adaptive scheme that can dynamically tune the MAC parameter settings for applications without strict latency requirements.

For IEEE 802.15.4 networks, parameters such as topology, node number, and length of data frame affect the performance of the network. Moreover, the CSMA/CA mechanism is employed in IEEE 802.15.4 when accesses to the channel lead to collisions and failure. Helena et al. showed that for a delivery rate of 99%, the maximum number of nodes in a star network should not exceed 32, and for a tree topology, the number of nodes must be less 18. Taking into account 5% hidden nodes, the maximum number of nodes in a star network must be less than 13 [9].

As far as we know, the previous studies concentrated on beacon networks [10–18] or simplified non-beacon networks [19–25]. However, the precision and practicability of these models to predict complex network behaviors were limited by many simplifications and assumptions. Therefore, it is necessary to model an IEEE 802.15.4 non-beacon network in a more complicated practical scenario, in which we consider a much more comprehensive complexity, e.g., the basic time unit, ACK frame, and retry limits. It is worth recalling that the relationships of these performance parameters are interdependent, but we always want to predict the network configuration when the several performance goals of the application have been identified and specified, e.g., we would like to predict the number of nodes and packet arrival rate when specifying the network performance parameter requirements. In consequence, we propose a useful prediction method to forecast parameters in practical applications that has not been considered in previous work. In our work, the non-beacon network with ACK mode and retransmission limit is modeled by three conditions: unsaturated traffic, saturated traffic, and MAC buffer size. Correspondingly, we propose three analytical models based on three-dimensional discrete time Markov models and an M/M/1/k queueing system. The basic time unit in our models is 1 symbol (16 μs), and all collisions are considered. The throughput, delay, and reliability performance metrics are obtained from these models. We developed a simulation program in C based on the Monte Carlo method. Furthermore, extensive simulations were used to validate the analytical results. Finally, backpropagation artificial neural networks (BP-ANNs) were used to predict the particular network parameters for a given quality of service.

The rest of the paper is organized as follows. Section 2 presents a wide variety of related models for the IEEE 802.15.4 network. In Section 3, we propose our analytical models and performance metrics of the IEEE 802.15.4 non-beacon network. The simulation results and analytical results are compared and discussed in Section 4. Section 5 concludes the paper.

## 2. Related work

There has been a significant amount of research that focuses on the performance metrics of beacon and non-beacon IEEE 802.15.4 networks, e.g., throughput, energy consumption, and delay. Some articles have mainly estimated the performance metrics by simulation or experimental tests [5–8]. Others have modeled the network using Markov models, renewal processes [27–28], or Petri Nets [29]. For beacon IEEE 802.15.4 networks, Ling [27] and Lee [28] used a renewal process to analyze the network. Lee [28] considered saturated and unsaturated traffic as well as ACK and non-ACK modes, and evaluated the throughput performance, average service time for a successful transmission, successful transmission probability, and dropping probability. Shuaib presented a deterministic Petri-net model of the IEEE 802.15.4 CSMA-CA process and extracted channel throughput and energy consumption [29]. Lee [28] indicated that Markov models were limited in their description of the unsaturated wireless sensor networks (WSNs) for nodes that do not always send packets. Most studies have employed the Markov model to analyze the beacon network. Sahoo modeled the ACK, retry limits, and unsaturated beacon networks using a three-dimensional discrete time Markov chain model. They analyzed the energy consumption and throughput [10]. Similarly, Park used a Markov chain, taking into account retry limits, ACKs, and unsaturated traffic [11]. The reliability, delay, and energy consumption under a low-traffic regime have also been obtained. Jung [12] considered a superframe structure, ACKs, and retransmissions with and without limits under unsaturated traffic conditions. They analyzed the throughput of the network.

Shu proposed a fixed-point formulation to investigate throughput performance under both saturation and non-saturation periodic traffic conditions, and considered both ACK and non-ACK modes [13]. Wang used a discrete time Markov model to analyze beacon and non-beacon networks [14]. Irfan used a Markov model and the M/G/1/L queue to analyze the end-to-end delay, reliability, and power consumption using different traffic and network conditions in star and cluster-tree WSN topologies [15]. Park [16] also established a Markov model for beacon networks using a derived simple expression at a certain accuracy. The model was used to derive a distributed adaptive algorithm for minimizing the power consumption within a given successful packet reception probability and delay constraints. Yin [17] combined the Geo/G/1 queueing model and Markov chain model to derive the throughput, delay and energy consumptions of beacon-enabled IEEE 802.15.4 network with unsaturated traffic. Wijetunge [18] evaluated the throughput of beacon-enabled IEEE 802.15.4 networks with hidden nodes.

Compared with beacon networks, non-beacon networks have higher throughput and probability of successful transmission [26]. Kim [19] used the Markov model and M/G/1 queueing system to analyze IEEE 802.15.4 unslotted CSMA/CA without ACK and retry limits, and obtained the throughput, delay, dropping probability, and energy consumption. Chiara [20] proposed a new model to analyze the non-beacon IEEE 802.15.4 network without ACK and retry limits. The model was able to obtain the distribution of network traffic, probability of a successful packet, and probability that a sink receives a packet.

Di Marco [21] presented a Markov chain model of the unslotted IEEE 802.15.4 MAC of multi-hop networks. ACK mode and retry limits were considered and only the probability of a successful packet was derived. Srivastava [22] developed a stochastic process for non-beacon IEEE 802.15.4 tree networks and obtained the reliability, delay, and throughput performance with and without hidden nodes. They assumed that ACKs seldom cause clear channel assessment (CCAs) to fail and rarely cause packet collisions. Moreover, their simulation results do not contain the ACK scenarios. Similarly, Feo [23] also analyzed non-beacon IEEE 802.15.4 multi-hop topologies with unsaturated traffic. They only derived the probability of channel access failure.

Wijetunge [24] set the discrete time unit to 1 symbol (16 μs) in a Markov chain model of a non-beacon IEEE 802.15.4 star network without ACK mode and retry limits. They derived the throughput of a network under unsaturated traffic. In summary, these models and simulations are less likely to perform well without the ACK mode and retry limits. Samaras [25] used Markov models to analyze non-beacon IEEE 802.15.4 single-hop hierarchical networks with ACK mode and retry limits in saturated and unsaturated conditions. The throughput and mean communication delay were derived.

However, it is still challenging to study network performance with respect to MAC buffer size with ACK mode and retry limits in practical scenarios. Therefore, our work differs from previous work [26] in the following respects. Our work strictly obeys the unslotted CSMA/CA specified in IEEE 802.15.4. The basic time unit in our models is 1 symbol (16 μs), and all collisions are considered. We use the Markov chain model to analyze IEEE 802.15.4 networks with ACK mode and retry limits. The models are used to describe the node and channel states. The MAC buffer size in the sensor node can range from one to an infinite number of packets, which can be modeled in the M/M/1/k queueing system. The unsaturated traffic used is also generated using a Poisson process in each sensor node. The performance of throughput, delay, and reliability are derived, and the simulation and analytical results are compared.

Artificial neural networks are a family of statistical learning models inspired by biological neural networks. These networks can learn patterns from the input data and, hence, can be used for classification. The interconnected neurons among every layer form the network. At present, researchers have used artificial neural networks to improve the performance of WSNs. Hwang used BP-ANNs to promote position accuracy [30]. Li presented an intruder detection system by combining fuzzy adaptive resonance theory neural networks and Markov models in WSN [31]. Nkwogu used four neural networks to predict the delay of data transmission in a wireless sensor actuator network (WSAN), and the BP-ANNs obtained a better performance [32]. Min proposed an improved handover algorithm using a BP-ANN in high-speed mobile WSN environments [33]. Veena used a hop field neural network to determine convergecast routes for wireless sensor networks [34].

Here, we used BP-ANNs to train the analytical data from the Markov chain model to predict three parameters separately, i.e., the node number *N*, successful transmission probability *PS*, and average successful transmission delay *TVS*. The comparison of the prediction results and the simulation results showed that this method was useful to help identify the limits of the *N*, *PS*, and *TVS* when deploying the WSN.

# 3 System model

The default parameters used in the analytical model for the unslotted CSMA/CA IEEE 802.15.4 network are given in Appendix A. We assume that there are $N$, ($N > 1$) sensor nodes and one sink node in the star topology network. All the nodes can hear each other, and there are no hidden nodes. The data packet length for every sensor node is $L$ bytes. The MAC buffer size is $M \times L$ bytes, where $M \geq 1$. The data frames arrive at the sensor nodes according to a Poisson arrival rate of $r$ frames per frame duration. The ACK frame and retransmission limits are considered in our models.

In this paper, we define the 1 mini-slot to be the minimum time unit in the unslotted CSMA/CA algorithm model. The duration of 1 mini-slot is 1 symbol, i.e., 16 μs. Thus, the discrete Markov model can be used to analyze the node state and channel state. In addition, the basic time unit in slotted CSMA/CA is 1 slot, which equals 20 mini-slots. In the unslotted CSMA/CA, we assume that all the nodes are synchronized to the same mini-slot and the data transmission begins at the boundary of the mini-slot. We assume the probability of sensing the channel at any mini-slot to be $\tau$, and every node senses the channel independently. The duration for sensing a node channel is 8 mini-slots. If the channel is idle over all 8 mini-slots, the CCA is idle. Otherwise, the CCA is busy when the channel is transmitting during any mini-slot of the 8 mini-slots.

Based on the description above, there are two possible data transmission collision, as follows.

(1) When the channel is idle with respect to CCA, if the difference in CCA start time between any two nodes in the network is equal or less than 12 symbols, the data frames from the two nodes will be in collision. Thus, the transmission fails.

(2) If the node starts a CCA and senses that the channel is idle when the channel is waiting for an ACK frame, the packet from the node will collide with the ACK frame, which leads to transmission failure.

In addition, the data frame will be discarded under three conditions. One is when the CCA failure number exceeds the *macMaxCSMABackoffs*, and another is when the retry failure number exceeds the retry limits. The last condition is when the node successfully sends a data packet and receives an ACK frame.

First, we propose the node state models under the three conditions in Section 3.1. Second, to describe the channel state clearly and obtain the performance parameters about the channel state, we also describe the behaviors of a channel using a Markov model in Section 3.2.

## 3.1 Node state models

We use a discrete time Markov model to describe the behaviors of a sensor node performing the unslotted CSMA/CA algorithm. Figs. 1–2 show the Markov models for different MAC buffer sizes under unsaturated traffic. Fig. 3 shows the Markov model under saturated traffic.

The node state in the backoff stage is defined as stochastic process $\{s(t), b(t), r(t)\}$, where $t$ is an integer and the unit is a min-slot. Further, *s(t)* is the backoff number at time t, where $s(t) \in [0, macMaxCSMABackoffs]$; *b(t)* is the backoff counter, where $b(t) \in [0, w_i - 1]$, $w_0 = 2^3 = 8$, $w_1 = 2^4 = 16$, $w_2 = 2^5 = 32$, $w_3 = 2^5 = 32$, and $w_4 = 2^5 = 32$; and *r(t)* is the retry counter, where $r(t) \in [0, aMaxFrameRetries]$. The duration of the $\{s(t), b(t), r(t)\}$ state is 20 symbols when *b(t)* > 0, and is 8 symbols when b(t) = 0.

Assume $\pi(s,b,r)=\lim_{t\to\infty}P\{s(t),b(t),r(t)\}$ is the stationary probability of $\{s(t),b(t),r(t)\}$. Here, $m$ denotes the retry counter, which is [0, *aMaxFrameRetries*].

The stationary probability of the idle state of the node is represented by $\pi(idle,m)$ when there are no packets to send. The duration of the $\{idle,m\}$ state is 1 symbol.

The stationary probability of the state of receiving to transmiting (RX-to-TX) turnaround time is represented by $\pi(TA,m)$. The duration of the $\{TA,m\}$ state is 12 symbols.

The stationary probability of the state when the node starts to send a packet for 12 symbols is represented by $\pi(TxC,m,n)$, where $n\in[0,12]$. The duration of the $\{TxC,m,n\}$ state is 1 symbol.

The stationary probability of the state of a successful transmission is represented by $\pi(TxSuc,m)$, and the duration of the $\{TxSuc,m\}$ state is $2L-12$ symbols.

The stationary probability of the state of the sensor node waiting for the ACK after sending the data packet is represented by $\pi(Turn,m)$. The duration of the $\{Turn,m\}$ state is 20 symbols.

The stationary probability of the state of the sensor node receiving the ACK frame at the first 12 symbols is represented by $\pi(AcW,m,n)$, where $n\in[1,12]$. The window in which the ACK may collide with the data packet is called the collision window. The duration of the $\{AcW,m,n\}$ state is 1 symbol.

The stationary probability of the state of the sensor node receiving the ACK frame in the last 10 symbols is represented by $\pi(AckSuc,m)$. The duration of the $\{AckSuc,m\}$ state is 10 symbols.

The stationary probability of a collision state of the sending data packet with another data packet or ACK frame is represented by $\pi(TF,m,n)$, where $n\in[0,12]$. The duration of the $\{TF,m,n\}$ state is 1 symbol, where $n\in[0,11]$. The duration of the $\{TF,m,12\}$ state is $2\times L-12$ symbols.

The stationary probability of the state of waiting for the ACK after sending the data packet is represented by $\pi(turn,m)$. The duration of the $\{turn,m\}$ state is 20 symbols.

The stationary probability of the state of the collision between the ACK and the data packet is represented by $\pi(AF,m,n)$. The duration of the $\{AF,m,n\}$ state is 1 symbol, where $n\in[0,11]$. The duration of the $\{AF,m,12\}$ state is 22 symbols. Here, the total duration of all $\{turn,m\}+\sum_{n=0}^{12}\{AF,m,n\}$ states is 54 symbols.

### 3.1.1 Model 1: MAC buffer size of *1 × L* bytes and unsaturated traffic

In Fig. 1, the MAC buffer size is *1 × L* bytes so that it can contain one packet. If the node is processing a data packet, any newly arriving data packets will be discarded. Assume that the data packet arrival rate is $r$ frames per data frame duration. The probability of a new data packet arrival is therefore $p = r/2L$. When the data packet has been successfully transmitted or discarded because of transmission failure, the sensor node will migrate to the idle state with a probability of 1. The probability of the sensor node in the idle state is $1 - p$. When a new data packet arrives with probability $p$, the sensor node will enter the first backoff stage. In the first backoff stage, the number of backoff periods ($NB$) is 0 and the backoff exponent ($BE$) is 3. The node will wait for a backoff period of a random number of units, allocated in uniform distribution. The range of the random value is in $[0, 2^{BE} - 1]$. After the delay, the node will turn to the {0,0,0} state for sensing the channel for 8 symbols. If the channel is busy, the node must go into the next backoff stage and the value of $NB$ and $BE$ increases by 1. When $NB$ exceeds 4, the node discards the data packet because five consecutive CCA failures have occurred.

If the node senses that the channel is idle, it will turn from the receiving state to the transmitting state after 12 symbols. The data packet is then sent to the sink node. The success and failure states are modeled in Fig. 1(b). The failure state can happen when there are at least two nodes transmitting the data packets or there is a transmission collision between an ACK frame and a data packet.

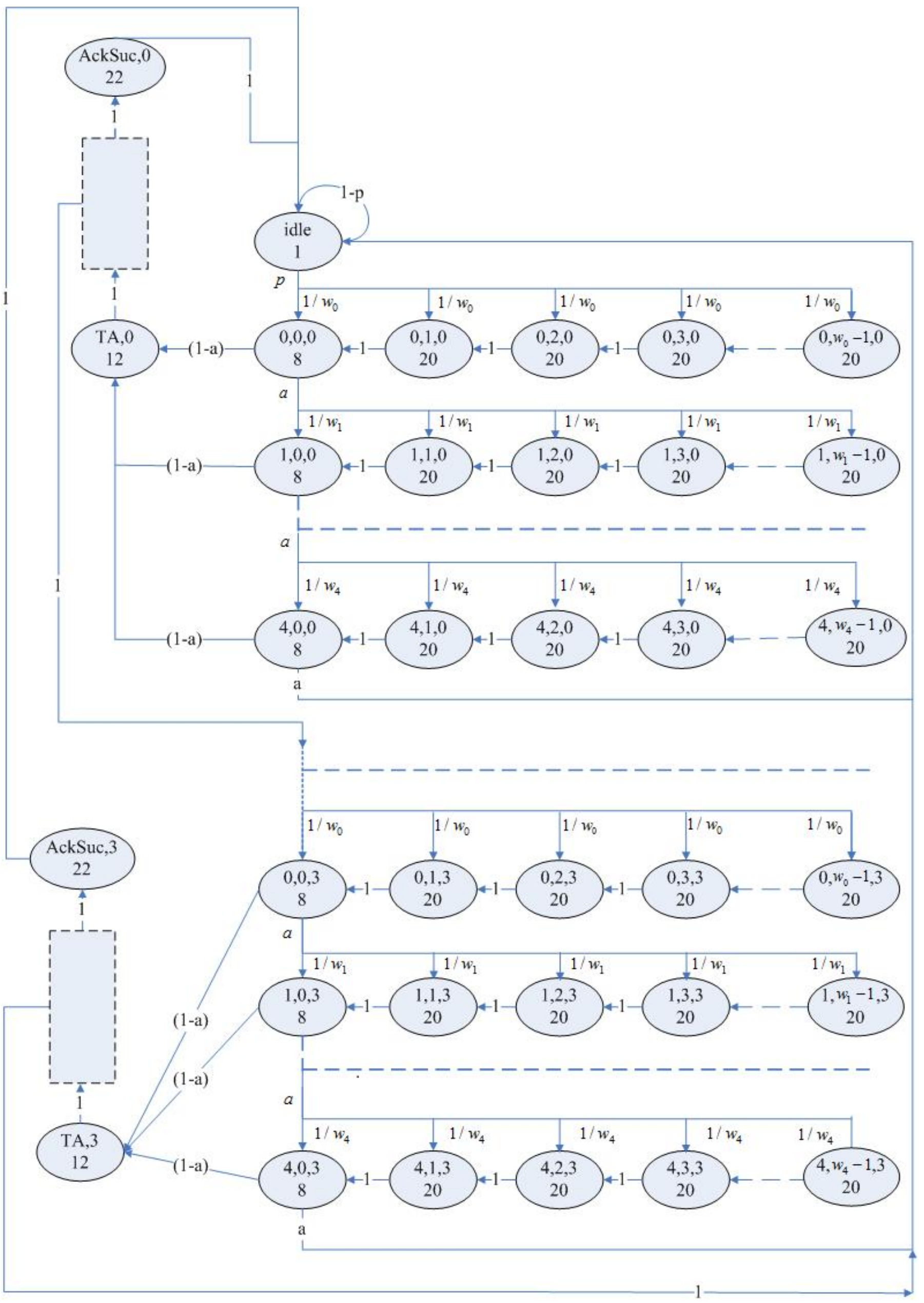

(a) Mac buffer = 1

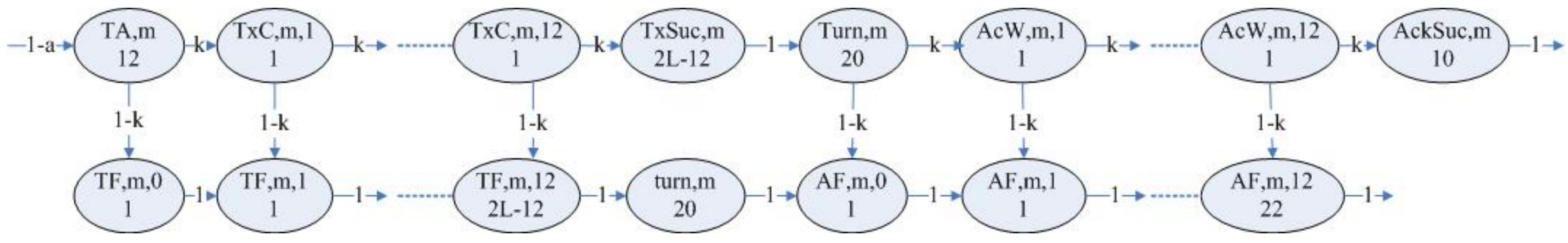

(b) All sub-states for the transmission state of the node that is represented in the dotted boxes in (a).

**Fig. 1** Markov chain for unslotted CSMA/CA with a 1 × L byte MAC buffer and unsaturated traffic. The number at the bottom of the oval represents the duration of the state.

In Fig. 1, *a* denotes the probability that a sensor node senses a busy channel when it starts a CCA.

The probability that the other $N-1$ sensor nodes do not start to sense the channel when the current node sends a packet is $k=(1-\tau)^{N-1}$.

Given the above description, we conclude that the transition probabilities in a steady state Markov chain model for the sensor node are as follows.

$$\pi(idle)=(1-p)\pi(idle)+\sum_{m=0}^{3}\pi(AckSuc,m)+\pi(AF,3,12)+a\sum_{m=0}^{3}\pi(4,0,m) \quad (1)$$

$$\pi(0,j,m)=\begin{cases}\frac{w_0-j}{w_0}p\pi(idle), & when \quad m=0;\\ \frac{w_0-j}{w_0}\pi(AF,m-1,12), & when \quad m>0;\end{cases} \quad (j\in[0,w_i],m\in[0,3]) \quad (2)$$

$$\pi(i,j,m)=\begin{cases}\frac{a}{w_i}\pi(i-1,0,m), when \quad j=w_i-1\\ \frac{a}{w_i}\pi(i-1,0,m)+\pi(i,j+1,m) \quad (i\in[1,4],j\in[0,w_i-2],m\in[0,3])\end{cases} \quad (3)$$

Transition probabilities for the other states are shown in Appendix B.

Let $\pi(idle,i)=\theta$ denote the probability of the steady state of a node in idle. To obtain $\tau$, we need to know the total mini-slots in all node steady states, which is denoted as $T$. First, assume $M1$ is the total number of mini-slots in node steady states when the retry counter is 0. This number can be expressed as follows.

$$\begin{aligned}M1&=20\sum_{i=0}^{4}\sum_{j=1}^{w_i-1}\pi(i,j,0)+8\sum_{i=0}^{4}\pi(i,0,0)+12\pi(TA,0)+\sum_{j=1}^{12}\pi(TxC0,j)+(2L-12)\pi(TxSuc,0)+20\pi(Turn,0)\\&+\sum_{j=1}^{12}\pi(AcW0,j)+10\pi(AckSuc,0)+\sum_{j=0}^{11}\pi(TF0,j)+(2L-12)\pi(TF0,12)+20\pi(turn,0)\\&+\sum_{j=0}^{11}\pi(AF0,j)+22\pi(AF0,12)\\&=[2L+144+158a+318a^2+318a^3+318a^4-12k^{26}-(20+66-12k^{26})a^5]\theta p\end{aligned} \quad (4)$$

It can be derived that $\pi(AF,0,12)=p\theta(1-a^5)(1-k^{26})$. We also assume that $D=(1-a^5)(1-k^{26})$, so the expression of $T$ can be simplified to

$$T = M1(1 + D + D^2 + D^3) + \theta \quad (5)$$

Based on these equations, the probability that the node starts a CCA can be expressed as follows.

$$\begin{aligned}\tau&=\frac{\sum_{i=0}^{3}\sum_{j=0}^{4}\pi(j,0,i)}{T}=\frac{(1+D+D^2+D^3)(1+a+a^2+a^3+a^4)\theta p}{M2(1+D+D^2+D^3)+\theta}\\&=\frac{(1+D+D^2+D^3)(1+a+a^2+a^3+a^4)}{[20L+144+158a+318a^2+318a^3+318a^4-12k^{26}-(20L+66-12k^{26})a^5](1+D+D^2+D^3)+\frac{1}{p}}\end{aligned} \quad (6)$$

## 3.1.2 Model 2: MAC buffer size is $M \times L$ bytes with unsaturated traffic

The Markov model of the node state is shown in Fig. 2 when the MAC buffer size is $M \times L$ ($M > 1$) bytes. The data packets arrive at the node following a Poisson process.

The data packets are processed in a FIFO sequence. The newly arrived data packets will be discarded if the MAC buffer is full. Based on queueing theory, we define $p = \lambda / \mu$ to be the utilization rate of the queue, where $\lambda$ is the arrival rate of the data packets, i.e., the average number of data packets arriving at the queue per unit time, and $\mu$ is the service rate of data packets, i.e., the average number of data packets leaving from the queue per unit time. When $p < 1$, the idle probability of the queue is $p0$. When $p \geq 1$, the queue is always in a busy state, i.e., the buffer is never empty. Therefore, the arrival rate of the data packet is $\lambda = r / 2L$. To obtain the average service time for every data packet, it is essential to consider the average time of successful transmissions and failed transmissions caused by exceeding retry limits or failing to access the channel. Let $TVS$ denotes the average service time. Then $\mu = 1 / TVS$.

Further, the utilization rate of the queue can be expressed as

$$p = \lambda / \mu = \frac{r \cdot TVS}{2L}. \quad (7)$$

Using M/M/1/k queueing theory, the probability that the MAC buffer becomes empty is

$$p0 = \frac{1 - p}{1 - p^{M+1}}. \quad (8)$$

Hence, the sensor node will migrate to the idle state with a probability of $p0$ or migrate to the first backoff stage with a probability $1 - p0$ when one data packet has finished transmitting.

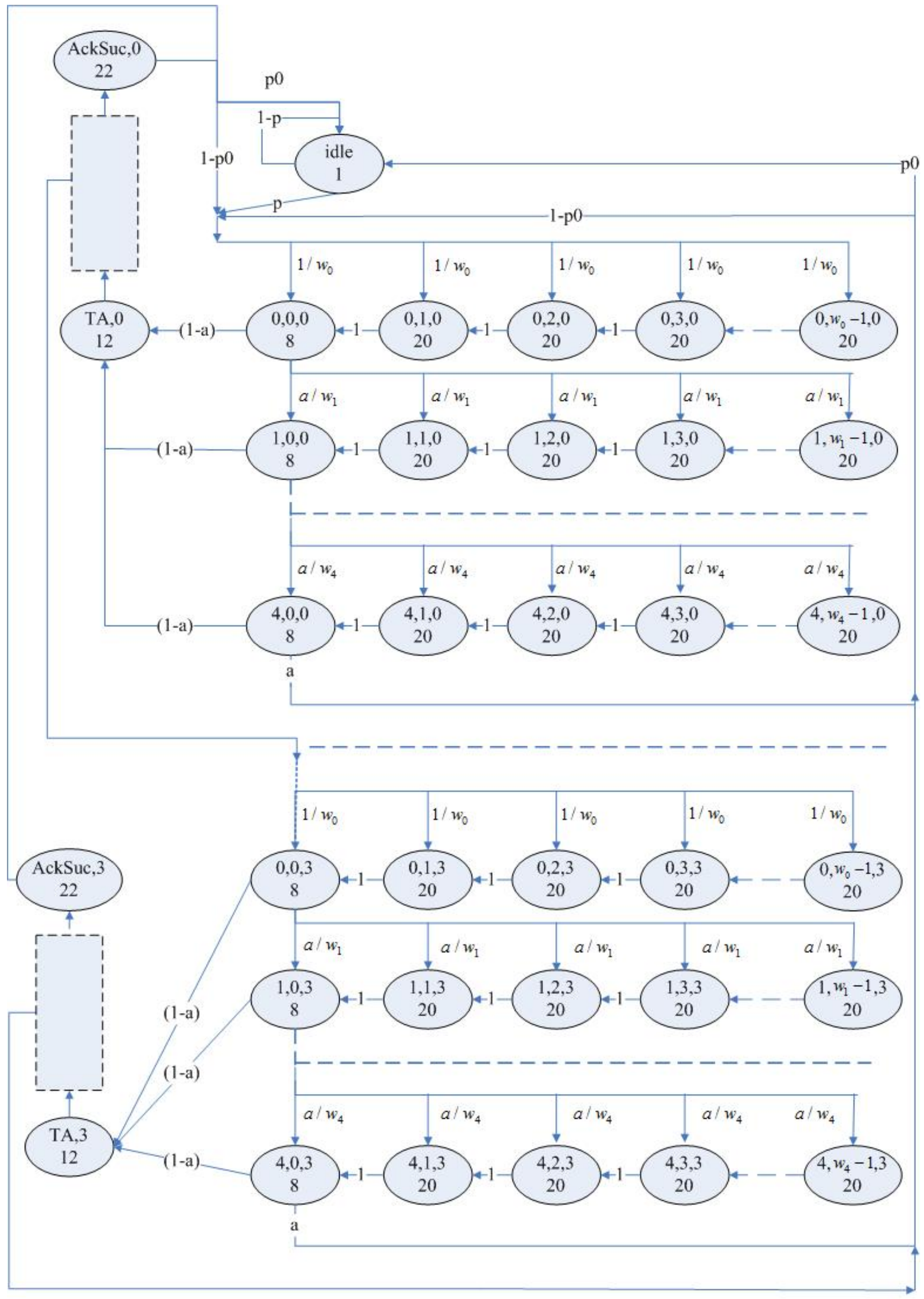

(a) Main state of the node

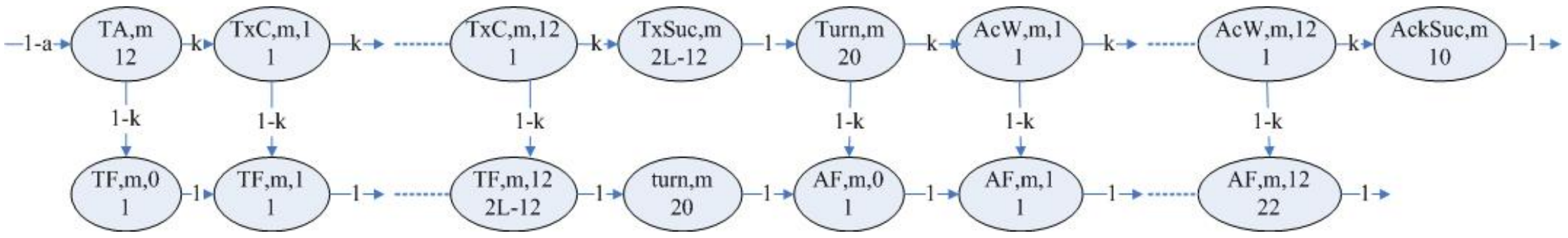

(b) All sub-states for the transmission state of the node represented in the dotted boxes in (a)

**Fig. 2** Markov chain for unslotted CSMA/CA with an M × L byte MAC buffer and unsaturated traffic. The number at the bottom of the oval represents the duration of the state.

The derivation of the formulas is very similar to those in Section 3.1.1. For simplicity, we only give the expression of the key parameters. The probability that the node starts a CCA in any time slot is

$$\tau = \frac{\sum_{i=0}^{3}\sum_{j=0}^{4}\pi(j,0,i)}{T} = \tag{9}$$
$$= \frac{(1+D+D^2+D^3)(1+a+a^2+a^3+a^4)}{[2L+144+158a+318a^2+318a^3+318a^4-12k^{26}-(2L+66-12k^{26})a^5](1+D+D^2+D^3)+\frac{p0\cdot 2L}{r}}$$

### 3.1.3 Model 3: Node state model with saturated traffic

We also consider the node state model with saturated traffic, as shown in Fig. 3. Under saturated traffic conditions, there is always a data frame waiting to be sent, and hence the MAC buffer is always busy. For simplicity, the probability that the node senses the channel without considering the data frame arrival rate is given by

$$\tau = \frac{\sum_{i=0}^{3}\sum_{j=0}^{4}\pi(j,0,i)}{T} = \frac{(1+D+D^2+D^3)(1+a+a^2+a^3+a^4)\theta p}{M2(1+D+D^2+D^3)+\theta} \tag{10}$$
$$= \frac{(1+D+D^2+D^3)(1+a+a^2+a^3+a^4)}{[2L+144+158a+318a^2+318a^3+318a^4-12k^{26}-(2L+66-12k^{26})a^5](1+D+D^2+D^3)}$$

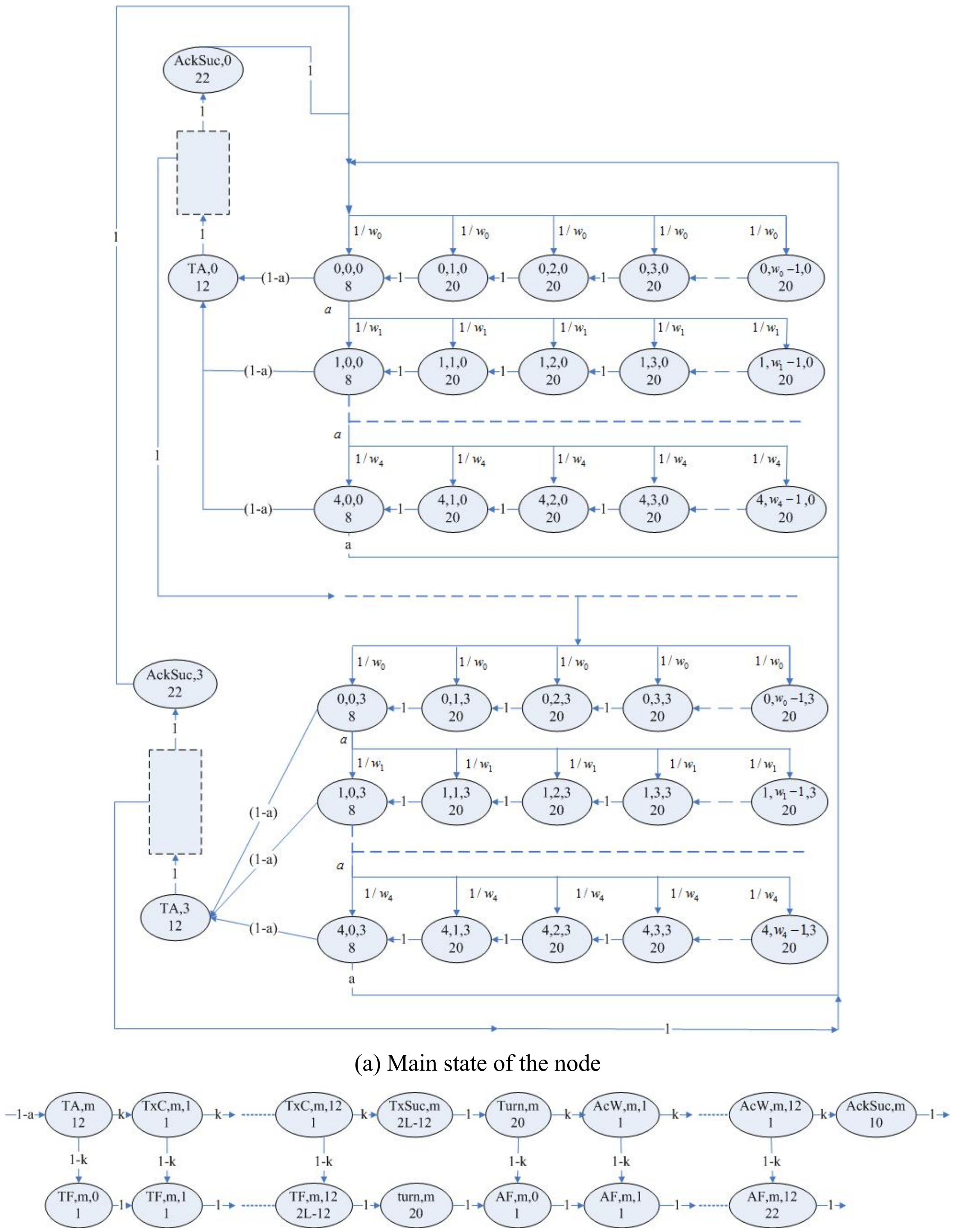

(b) All sub-states for the transmission state of the node represented in the dotted boxes in (a)

**Fig. 3** Markov chain for unslotted CSMA/CA with saturated traffic. The number at the bottom of the oval represents the duration of the state.

## 3.2 Channel state model

To calculate the performance parameters of the channel such as throughput and the probability that the channel busy, we also use the discrete time Markov chain to model the state of the channel. The ACK and retry limits are considered in this channel model.

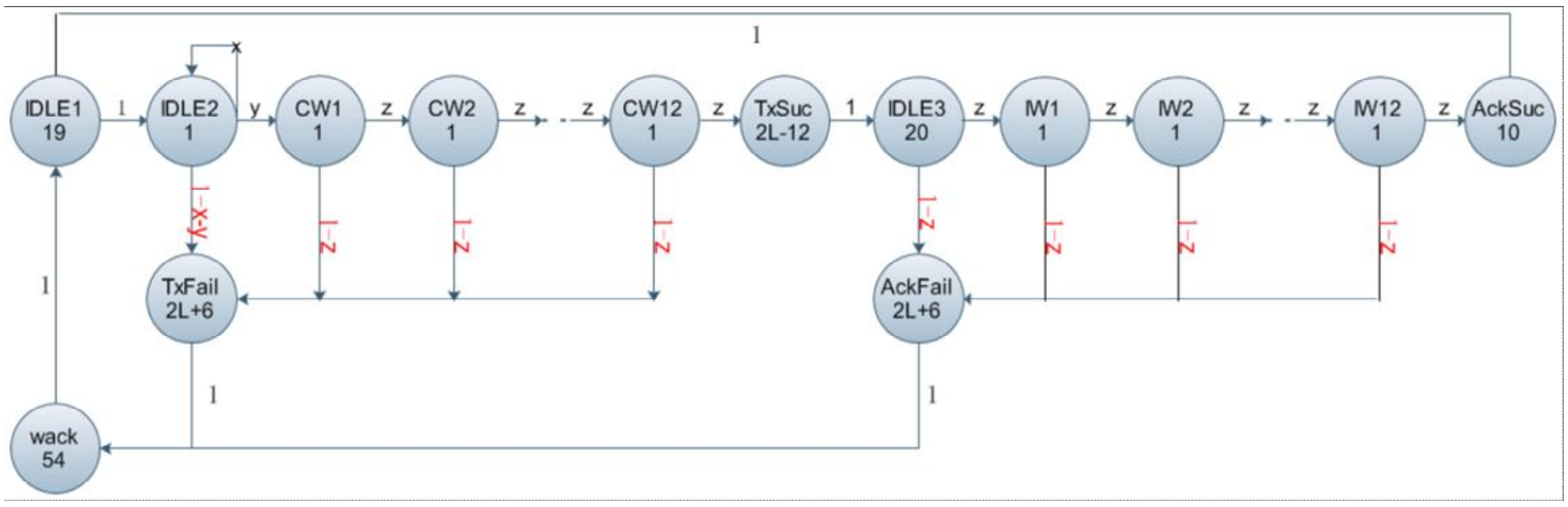

**Fig. 4** Model of the channel state. The number at the bottom of the circle represents the duration of the state.

In Fig. 4, $x$ denotes the probability that none of the sensor nodes sense the channel, then

$$x = (1-\tau)^N . \quad (11)$$

Further, $y$ denotes the probability that only one sensor node is transmitting or sensing the channel at the given mini-slot (= 1 symbol), that is,

$$y = N\tau(1-\tau)^{N-1} . \quad (12)$$

Finally, $z$ denotes the probability that none of the remaining $N - 1$ sensor nodes transmit the data packet:

$$z = (1-\tau)^{N-1} . \quad (13)$$

The time taken for the sensor node to send the data packet from the backoff state is at least 20 symbols. Assume that {IDLE1} denotes the state of first 19 symbols after the sensor node starts backoff. State {IDLE2} represents the state before sending the data packet; and we set its time to 1 symbol. The sensor node may be in this state with a probability of $x$.

State $\{CW_i\}(i \in [1,12])$ is one of the states of the first 12 symbols when the channel starts transmitting the data packet. The first 12 states may be in collision with other data packets or ACK frames. If the collision happens, the channel will migrate to the {TxFail} state with a probability of $1 - z$.

State {TxSuc} is the state of the time remaining for the successful transmission of a data packet. The duration of {TxSuc} is $2L - 12$ symbols.

State {IDLE3} is the state of the turnaround time from the *TX* state to the *RX* state, and its duration is 20 symbols.

State {AckSuc} is the state of the time remaining for the successful transmission of the ACK frame.

The duration of *AckSuc* is 10 symbols.

State $\{IW_i\}(i \in [1,12])$ is the state of every symbol in the first 12 symbols when the channel starts transmitting the ACK frame. In this state, there may be a collision with other data packets or ACK frames. The channel state then enters the {AckFail} state.

State {TxFail} is the state of the failed transmission of the data packet. Its duration is $2 \times L + 6$ symbols on average.

State {AckFail} is the state of the failed transmission of the ACK frame. Its duration is $2 \times L+6$ symbols on average.

State {wack} is the state of waiting for the ACK frame after a failed data packet transmission. Its duration is 54 symbols.

Let $\pi(IDLE2) = \theta$ denote the stationary probability of the {IDLE2} state. We can then derive the stationary probability of the other states in the channel model, which are given in Appendix C.

Define T as

$$\begin{aligned} T &= 19\pi(IDLE1) + \pi(IDLE2) + \sum_{i=1}^{12}\pi(CWi) + (2L-12)\pi(TxSuc) + 20\pi(IDLE3) \\ &+ \sum_{i=1}^{12}\pi(IWi) + 10\pi(AckSuc) + (2L+6)\pi(TxFail) + (2L+6)\pi(AckFail) + 54\pi(wack) \\ &= \theta \cdot [2L + 80 - (2L+79)\cdot x + y\sum_{i=0}^{24} z^i + (2L+7)\cdot y \cdot z^{12} - (2L+50)\cdot y \cdot z^{25}] \end{aligned} \quad (14)$$

The probability that the sensor node senses the channel is idle can be expressed as follows.

$$\begin{aligned} 1-a &= \frac{\frac{12}{19}\cdot 19\pi(IDLE1) + \pi(IDLE2) + \frac{13}{20}\cdot 20\pi(IDLE3)}{T} \\ &= \frac{12(1-x)+1}{2L + 80 - (2L+79)\cdot x + y\sum_{i=0}^{24} z^i + (2L+7)\cdot y \cdot z^{12} - (2L+50)\cdot y \cdot z^{25}]} \end{aligned} \quad (15)$$

Using the above equations, we can derive the normalized throughput *TH* as

$$\begin{aligned} TH &= \frac{2L\pi(AckSuc)}{T} \\ &= \frac{2L\cdot y \cdot z^{25}}{2L + 80 - (2L+79)\cdot x + y\sum_{i=0}^{24} z^i + (2L+7)\cdot y \cdot z^{12} - (2L+50)\cdot y \cdot z^{25}]} \end{aligned} \quad (16)$$

## 3.3 Performance parameters

In Sections 3.1–3.2, we propose three node state models, and in Section 3.3, we propose a common channel model. For every model, we can combine the node state model with the channel model to derive the performance metrics. The values of $\tau$ and $a$ can be solved by numerical methods. For unsaturated traffic, when the MAC buffer size is $1 \times L$, $\tau$, $a$, and $TH$ can be derived from formulas (6) and (15)–(16). In addition, when the MAC buffer size is $M \times L$, $\tau$, $a$, and $TH$ can be derived from formulas (7)–(9) and (15)–(16). For saturated traffic, $\tau$, $a$, and $TH$ can be derived from formulas (10) and (15)–(16). Moreover, the other parameters in Table 1 can be derived from $\tau$ and $a$.

We derived eight performance parameters from the above three Markov models with unsaturated and saturated traffic. Table 1 lists these parameters.

Table 1 Eight performance parameters

| Unsaturated traffic M=1 | Unsaturated traffic M>1 | Saturated traffic | Meaning |
|---|---|---|---|
| τ | τ | τ | The probability of sensing the channel |
| a | a | a | The probability of a CCA failure due to the channel is busy |
| TH | TH | TH | The normalized throughput |
| PS | PS | PS | The probability of successful transmission of data frame |
| TS | TS | TS | The delay of the successful data transmission (without queueing delay) |
| TVS | TVS | TVS | The average delay of the data transmission without queueing delay |
| | TSW | | The delay of the successful data transmission with queueing delay |
| | TVSW | | The average delay of the data transmission with queueing delay |

The other parameters can be derived as follows. Assume T1 to be the average time of the data packet deleted from the MAC buffer when the number of the sensor node fails to access the channel because *macMaxCSMABackoffs* has been exceeded. Then,

$$T1 = \sum_{i=0}^{4}(b_i + t_{CCA}) = 1190, \quad (17)$$

where $b_i = \frac{(w_i - 1)*20}{2}, b_0 = 70, b_1 = 150, b_2 = 310, b_4 = 310, b_4 = 310$.

Assume T2 is the average time for the sensor node to successfully send the data packet and receive an ACK frame in one transmission. Then,

$$\begin{aligned} T2 = {} & \frac{1}{(1-a^5)k^{26}}[(1-a)k^{26}(b_0 + t_{CCA} + ta + 2L + turn + tack) + \\ & a(1-a)k^{26}(b_0 + b_1 + 2t_{CCA} + ta + 2L + turn + tack) + \\ & a^2(1-a)k^{26}(b_0 + b_1 + b_2 + 3t_{CCA} + ta + 2L + turn + tack) + \\ & a^3(1-a)k^{26}(b_0 + b_1 + b_2 + b_3 + 4t_{CCA} + ta + 2L + turn + tack) + \\ & a^4(1-a)k^{26}(b_0 + b_1 + b_2 + b_3 + 4t_{CCA} + ta + 2L + turn + tack)] \\ = {} & \frac{1}{1-a^5}[132 + 158a + 318a^2 + 318a^3 + 318a^4 - 1244a^5 + 2L(1-a^5)] \end{aligned}, \quad (18)$$

where $t_{CCA}$ is the duration of CCA (8 symbols), *ta* is the duration of the RX-to-TX turnaround time (12 symbols), *turn* is the duration of the sensor node waiting for the ACK frame after sending the data packet (20 symbols), and *tack* is the duration of the ACK frame (22 symbols).

Assume T3 is the average time for a data packet to collide with other data packets or the ACK frame in one transmission. We obtain

$$\begin{aligned} T3 = {} & \frac{1}{(1-a^5)(1-k^{26})}[(1-a)(1-k^{26})(b_0 + t_{CCA} + ta + 2L + wack) + \\ & a(1-a)(1-k^{26})(b_0 + b_1 + 2t_{CCA} + ta + 2L + wack) + \\ & a^2(1-a)(1-k^{26})(b_0 + b_1 + b_2 + 3t_{CCA} + ta + 2L + wack) + \\ & a^3(1-a)(1-k^{26})(b_0 + b_1 + b_2 + b_3 + 4t_{CCA} + ta + 2L + wack) + \\ & a^4(1-a)(1-k^{26})(b_0 + b_1 + b_2 + b_3 + 4t_{CCA} + ta + 2L + wack)] \\ = {} & \frac{1}{1-a^5}[144 + 170a + 330a^2 + 330a^3 + 330a^4 - 1256a^5 + 2L(1-a^5)] \end{aligned}, \quad (19)$$

where *wack* indicates the duration of waiting for the ACK frame, which is 54 symbols.

In each transmission, we assume that the probability of a successful transmission for a node is *PSuc*, the probability of data transmission failure due to CCA failure is *PAcc*, and the probability of a collision transmission for a node is *PColl*. Because the probability of the node starts the CCA is $\tau$, the expressions of *PSuc*, *PAcc*, and *PColl* are as follows.

$$PSuc = \frac{\tau \cdot (1-a^5)k^{26}}{\tau \cdot (1-a^5)k^{26} + \tau \cdot a^5 + \tau \cdot (1-a^5)(1-k^{26})} \quad (20)$$

$$PAcc = \frac{\tau \cdot a^5}{\tau \cdot (1-a^5)k^{26} + \tau \cdot a^5 + \tau \cdot (1-a^5)(1-k^{26})} \quad (21)$$

$$PColl = \frac{\tau \cdot (1-a^5)(1-k^{26})}{\tau \cdot (1-a^5)k^{26} + \tau \cdot a^5 + \tau \cdot (1-a^5)(1-k^{26})} \quad (22)$$

Let $PS_i$ be the probability of a successful data transmission when the retransmission counter is i, where i ∈ [0, 3]. The probability that the other *N − 1* sensor nodes do not start to sense the channel when the current node sends a packet is $k=(1-\tau)^{N-1}$.

Let $PC_i$ be the probability of data transmission failure due to CCA failure when the retransmission counter is i, where i ∈ [0, 3]. Further, let $PF_i$ be the probability of data retransmission failure when the retransmission counter is i, where i ∈ [0, 3].

When the retransmission counter is 0, the expressions of $PS_0$, $PC_0$, and $PF_0$ are as follows.

$$PS_0=(1-PColl-PColl^2-PColl^3)PSuc \quad (23)$$

$$PC_0=(1-PColl-PColl^2-PColl^3)PAcc \quad (24)$$

$$PF_0=PColl \quad (25)$$

When the retransmission counter is greater than 0, the expressions of $PS_i$, $PC_i$, and $PF_i$ (i > 0) are as follows.

$$PS_i=PColl^iPSuc \qquad (i=1..3) \quad (26)$$

$$PC_i=PColl^iPAcc \qquad (i=[1,3]) \quad (27)$$

$$PF_i=PColl^{i+1} \qquad (i=[1,3]) \quad (28)$$

We can obtain the probability of successful data transmission for a node when the maximum number of retry times is three.

$$PS=\sum_{i=0}^{3}PS_i/(\sum_{i=0}^{3}PS_i+\sum_{i=0}^{3}PC_i+PF_3) \quad (29)$$

We can then calculate the delay of a successful data transmission. The expression is as follows.

$$TS=\frac{\sum_{i=0}^{3}PS_i[i\cdot T3+T2]}{PS} \quad (30)$$

The average delay of data transmission *TVS* is as follows.

$$TVS=(\sum_{i=0}^{3}PC_i[i\cdot T3+T1]+\sum_{i=0}^{3}PS_i[i\cdot T3+T2]+4\cdot PF_3\cdot T3)/(\sum_{i=0}^{3}PC_i+\sum_{i=0}^{3}PS_i+PF_3) \quad (31)$$

We need to consider the queueing delay when the MAC buffer can contain one more data frame. Based on M/M/1/k queueing theory, let $L_q$ be average number of data frames waiting to be sent in the node. It can be expressed as

$$L_q = \frac{p}{1-p} - \frac{(M+1)p^{M+1}}{1-p^{M+1}} - (1-p_0) = \frac{p}{1-p} - \frac{(M+1)p^{M+1}}{1-p^{M+1}} - \frac{(1-p^M)p}{1-p^{M+1}} \quad (p \neq 1), \ (32)$$

where $p$ is the utilization rate of the queue, as detailed in Section 3.1.1, and $M$ is the maximum number of data frames allowed in the MAC buffer.

Assume $\lambda_e$ to be the effective arrival rate in the queue, expressed as

$$\lambda_e = \lambda(1-p_M), \ (33)$$

where the arrival rate of a data packet is $\lambda = r / 2L$, and the probability that the MAC buffer is full is $p_M = \frac{(1-p)p^M}{1-p^{M+1}}$.

Based on Little's law from queueing theory, the average waiting time for every data frame is

$$W_q = \frac{L_q}{\lambda_e}. \ (34)$$

Therefore, when considering the queuing time, we can derive the expressions of TSW and TVSW, respectively, as follows.

$$\text{TSW} = \text{TS} + \text{W}_\text{q} \ (35)$$

$$\text{TVSW} = \text{TVS} + \text{W}_\text{q} \ (36)$$

# 3.4 BP-ANN model

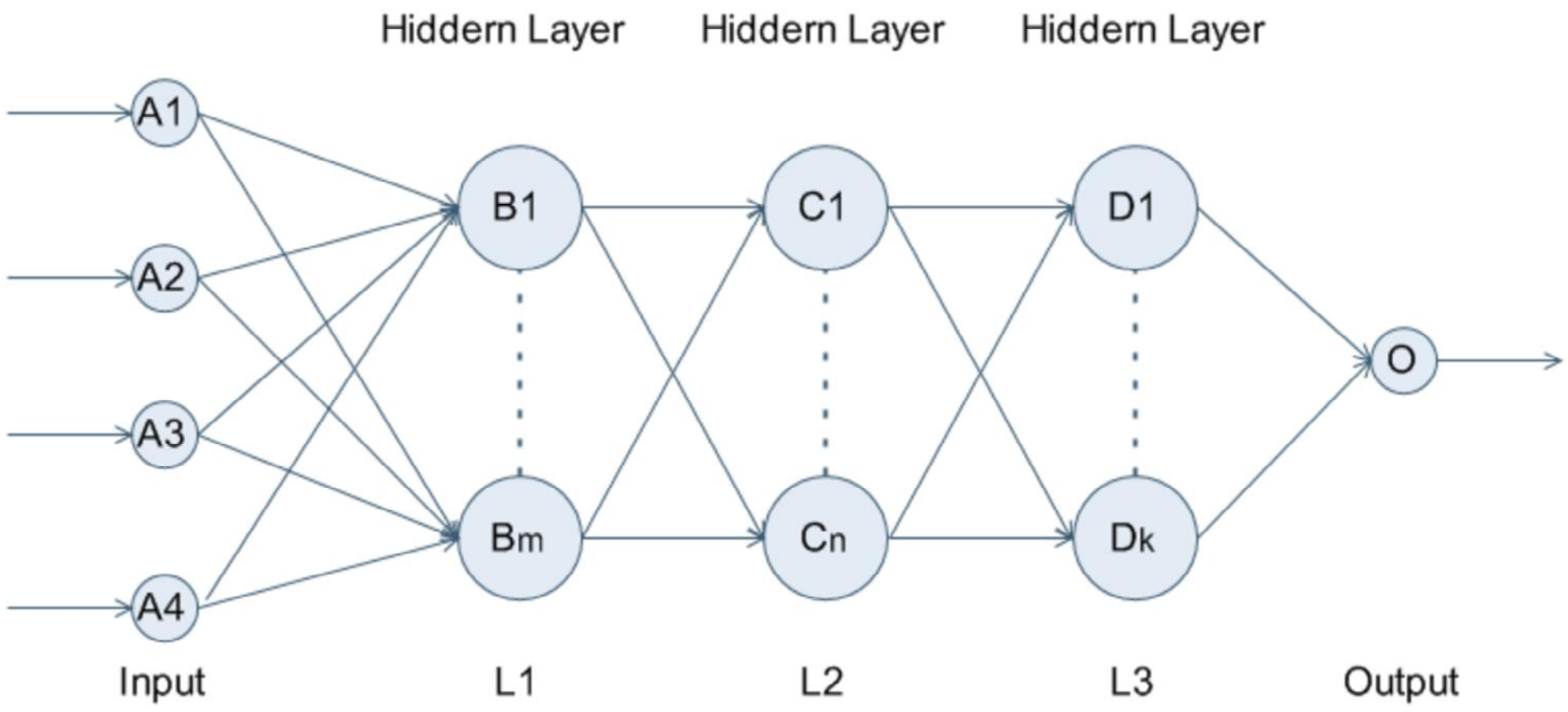

**Fig. 5** BP-ANN model. The input layer has four variables, three hidden layers with sigmoid neurons, and one linear output layer. L1, L2, and L3 are the numbers of nodes in every hidden layer.

In the present work, the BP-ANN algorithm is used to build a nonlinear prediction model for determining wireless sensor network configuration and specific performance parameters to meet the given requirements. In this BP-ANN model, there are tree hidden layers connecting the input and output layers and the activation function is a sigmoid-type function. The organization of the BP-ANN is shown in Fig. 5.

The nodes of the input layer depend on input parameters, e.g., the number of sensor nodes, the number of successful data frame transmissions, and average transmission delay of data frames. We built four BP-ANN models to predict different performance parameters, as illustrated in Table 2.

Table 2 BP-ANN models

| **Model name** | **Training set** | **Testing set** | **The input vector** | **The three hidden layers [L1, L2, L3]** | **Output** |
|---|---|---|---|---|---|
| **Model1** | **60,000** | 100 | [r,L,PS,TVS] | [100,80,50] | N |
| **Model2** | **60,000** | 100 | [r,L,N,TVS] | [80,50,30] | PS |
| **Model3** | **60,000** | 100 | [r,L,PS,N] | [80,50,30] | TVS |

We used four given parameters under specific conditions to predict the fifth parameter. The data frame length is $L$, where $L \in [30, 127]$ bytes. The data frame arrival rate is $r$, where $r \in [0.001, 0.1]$ frames/frame duration. The probability of successful data frame transmission is $PS$, where $PS \in [0, 1]$. The average data frame transmission delay is $TVS$, where the time unit is 1 symbol. Finally, the number of nodes is $N$, where $N \in [2,100]$.

## 4. Results and discussion

To validate the Markov chain model of IEEE 802.15.4 network, we developed a simulation program in C based on Monte Carlo methods for the non-beacon enabled star network. In the star network, there are $N$ wireless sensor nodes and one sink node. The sensor node has a MAC buffer. We assume that all data frames have the same length and the data frames arrive at the MAC buffer with a Poisson process. In each simulation condition, the number of iterations is 50. We do not consider hidden nodes.

Assume the length of the data frame is $L$ bytes, the size of MAC buffer is $M \times L$ bytes ($M \geq 1$), the arrival rate of the data frame is $r$ (frames/frame duration), where the frame duration is $2 \times L$ symbols and one symbol duration is 16 μs. The unit of r is frames/(1000 × frame duration), and the frame duration is $2 \times L$ symbols.

## 4.1 Probability of CCAs and sensing a busy channel

With unsaturated traffic, according to Equations (6), (9), and (10), the probability of a CCA for the node is related to the node number, packet length, packet arrival rate, the probability of sensing a busy channel, and the MAC buffer size. With saturated traffic, according to Equation (10), the probability of a CCA for the node is related to the node number, packet length, and the probability of sensing a busy channel. The influence of node number $N$ on the CCA probability is shown in Fig. 6(a). Here, the MAC buffer size is $1 \times L$ bytes. Parameter $\tau$ increases when $r$ and $N$ increase. Because the increase of node number and arrival rate leads to more data frame transmissions, the CCA probability also increases. The influence of packet length $L$ on the CCA probability is shown in Fig. 6(b). Here, the MAC buffer size is $1 \times L$ bytes. The figure shows that $\tau$ decrease as $L$ increases. This is because an increase in $L$ will lead to a decrease in collisions and an increase in packet transmission delay. The influence of MAC buffer size on the CCA probability is shown in Fig. 6(c), where $L = 100$ bytes and $N = 10$. This figure shows that τ increases as $M$ increases. A larger MAC buffer size can contain more packets that immediately start a CCA after the last packet transmission; hence, the frequency of CCAs increases. The influence of $N$ and $L$ on the CCA probability with saturated traffic is shown in Fig. 6(d). This figure shows that $\tau$ decreases as $L$ increases. The reason for this is similar to that shown in Fig. 6(b).

With unsaturated traffic, $\tau$ will steadily increase as r, N, and M increase, as shown in Fig. 6(c). In contrast, it decreases as L increases. With saturated traffic, $\tau$ is almost steady when $N > 5$. This indicates that the nodes are always in a busy state.

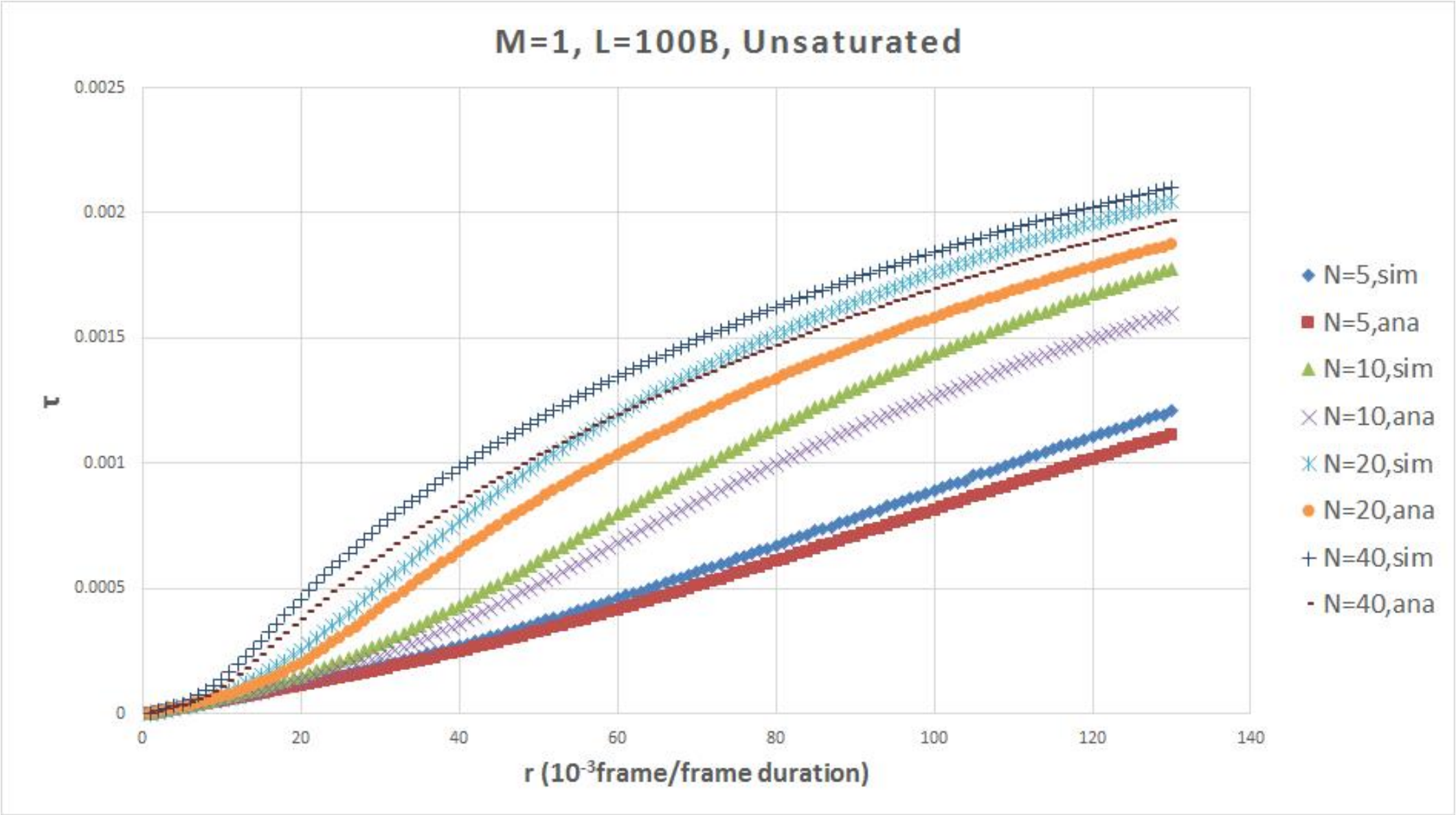

(a) Changes in τ for different node numbers

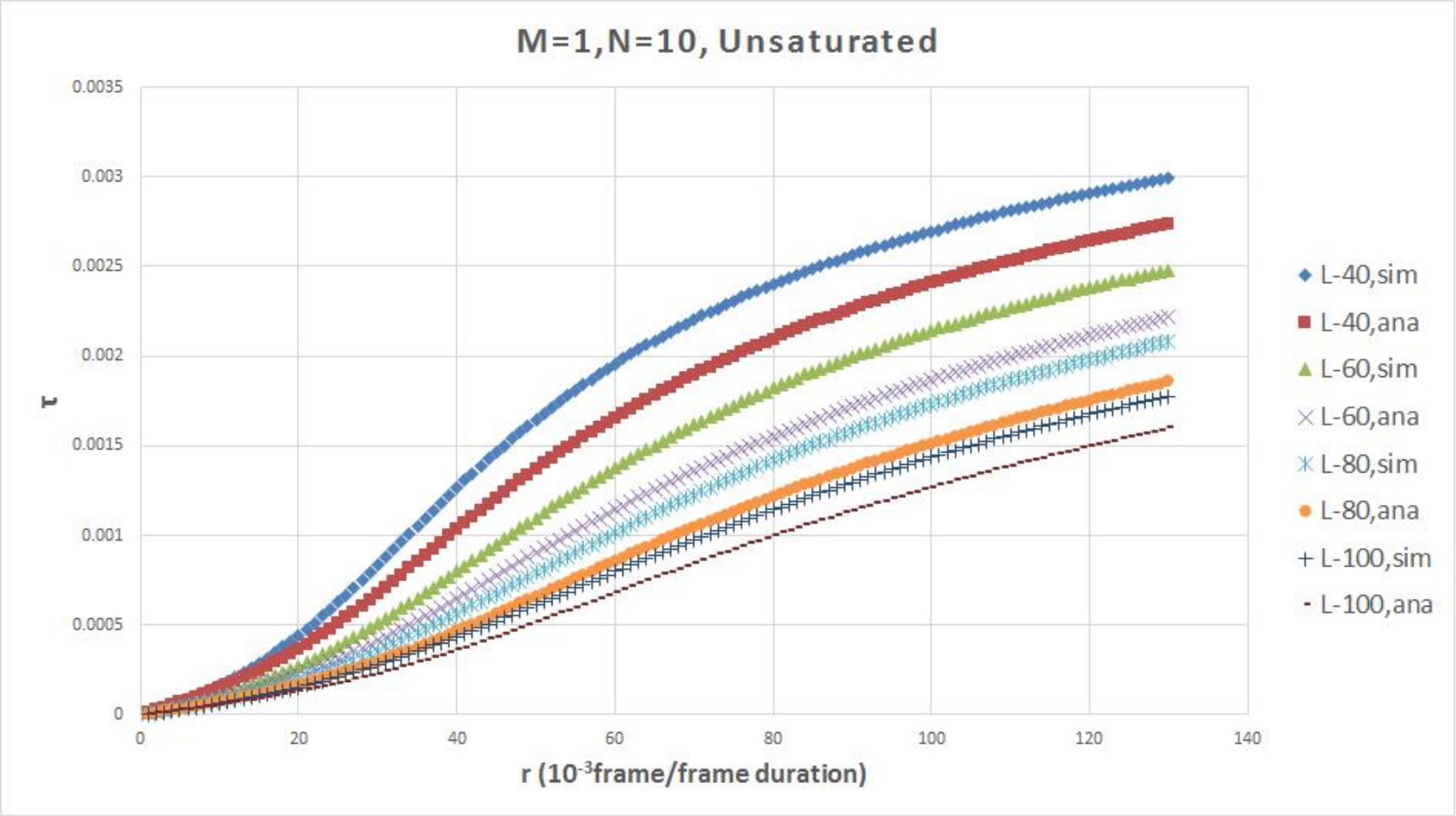

(b) Changes in τ for different frame lengths

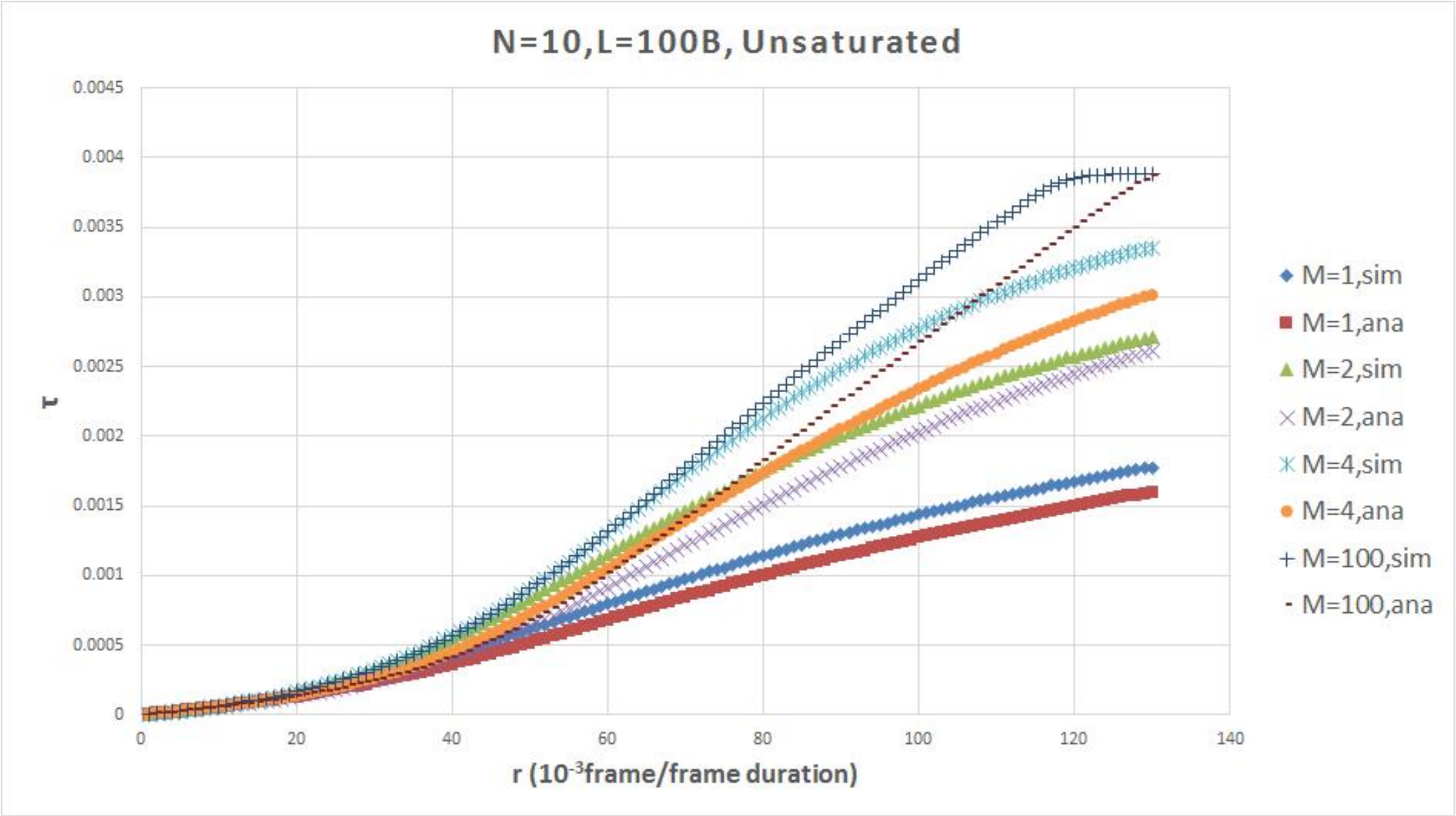

(c) Changes in τ for different buffer sizes

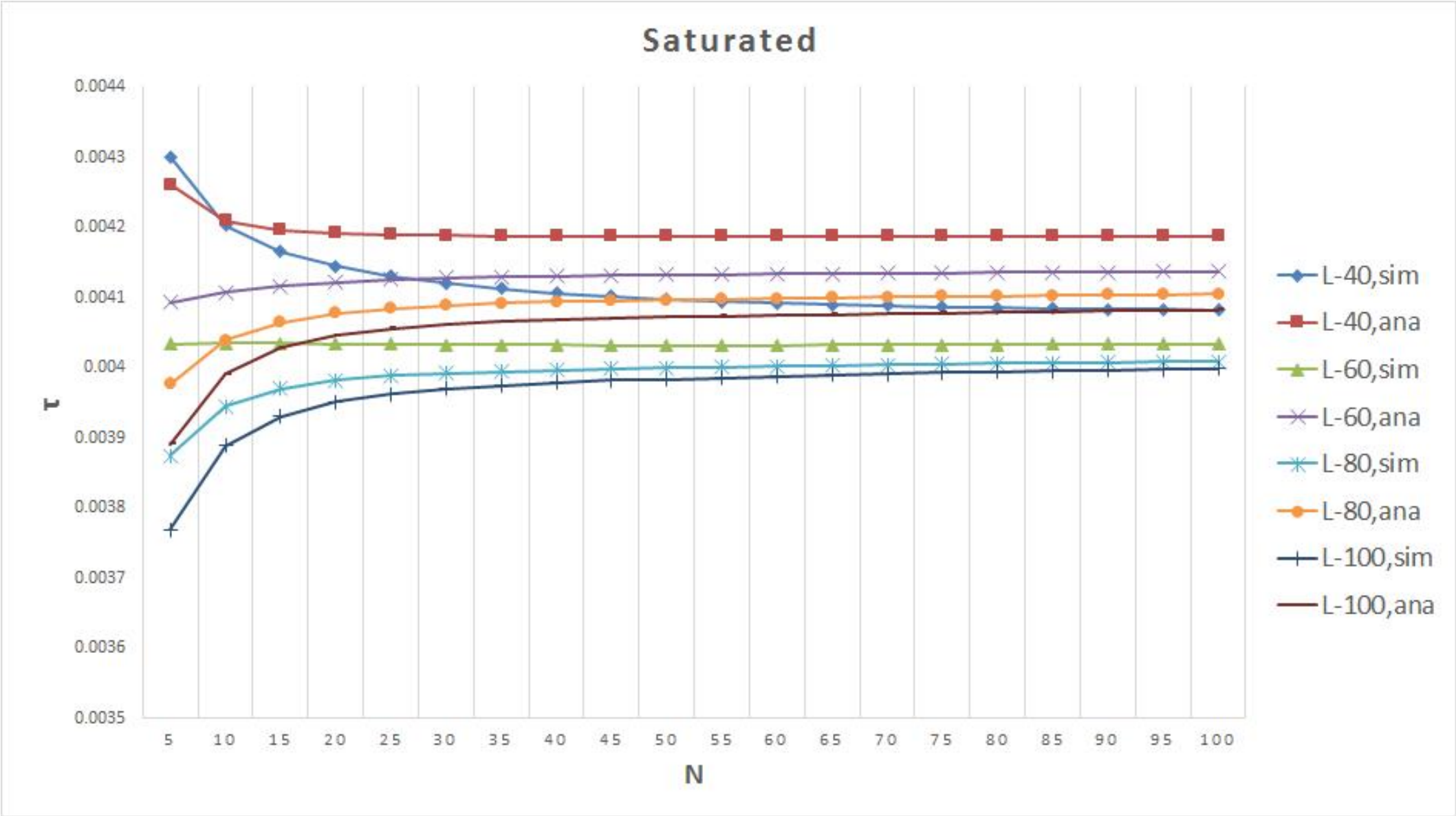

(d) Changes in τ for different frame lengths in the saturated case

**Fig. 6** Probability that the node starts a CCA in a random mini-slot

According to Equation (15), the probability of sensing a busy channel is related to the node number, packet length, and CCA probability. The influence of node number $N$ on the probability of sensing a busy channel is illustrated in Fig. 7(a), where $L$ = 100 bytes, $r \in [0.001, 0.13]$, and M = 1. The influence of packet length L on the probability $a$ of sensing a busy channel is illustrated in Fig. 7(b), where $L$ = 100 bytes, $r \in [0.001, 0.13]$, and $M = 1$. When the arrival rate is held fixed, $a$ decreases as $L$ increases. This is because larger packet length will reduce the collision probability. The influence of the MAC buffer size on the probability of sensing a busy channel is illustrated in Fig. 7(c), where $L$ = 100 bytes, $r \in [0.001, 0.13]$, and $N = 10$. The figure shows that $a$ increases as $M$ increases.

As shown in Fig. 7(d), both simulation and analytical results indicate the influence of $N$ and $L$ on the probability of sensing a busy channel with saturated traffic. This figure shows that $a$ will decrease as $L$ increases.

In Fig. 7, $a$ increases as r, N, and M increase, whereas it decreases as L increases when the network becomes saturated. With unsaturated traffic, $a$ can obtain a steady state quickly when N or r increases. With saturated traffic, $a$ is almost in a steady state when N > 20. These results show that the non-beacon star network can only handle a small number of nodes or low rate of transmission if a low CCA failure is required.

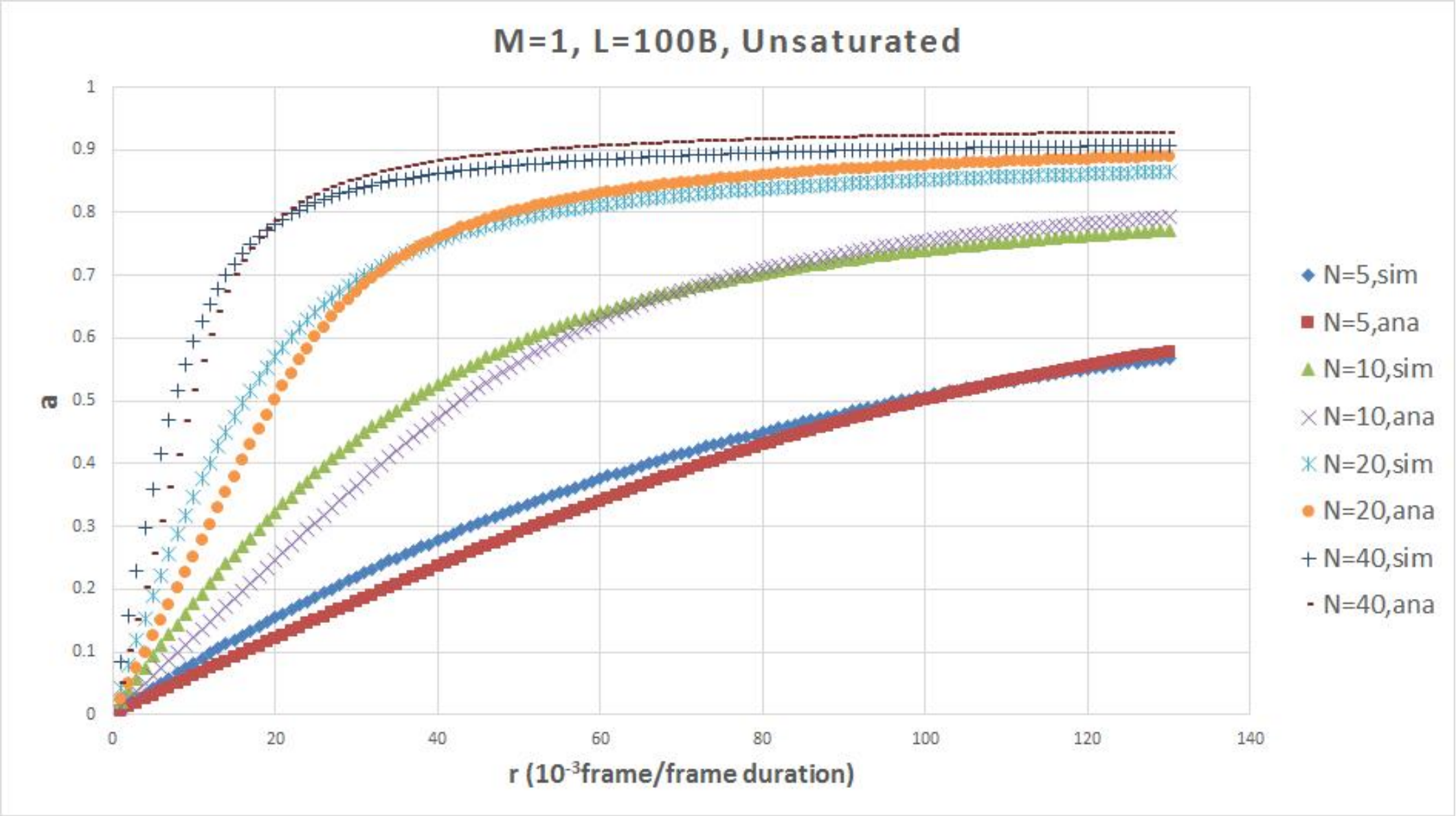

(a) Changes in $a$ for different node numbers

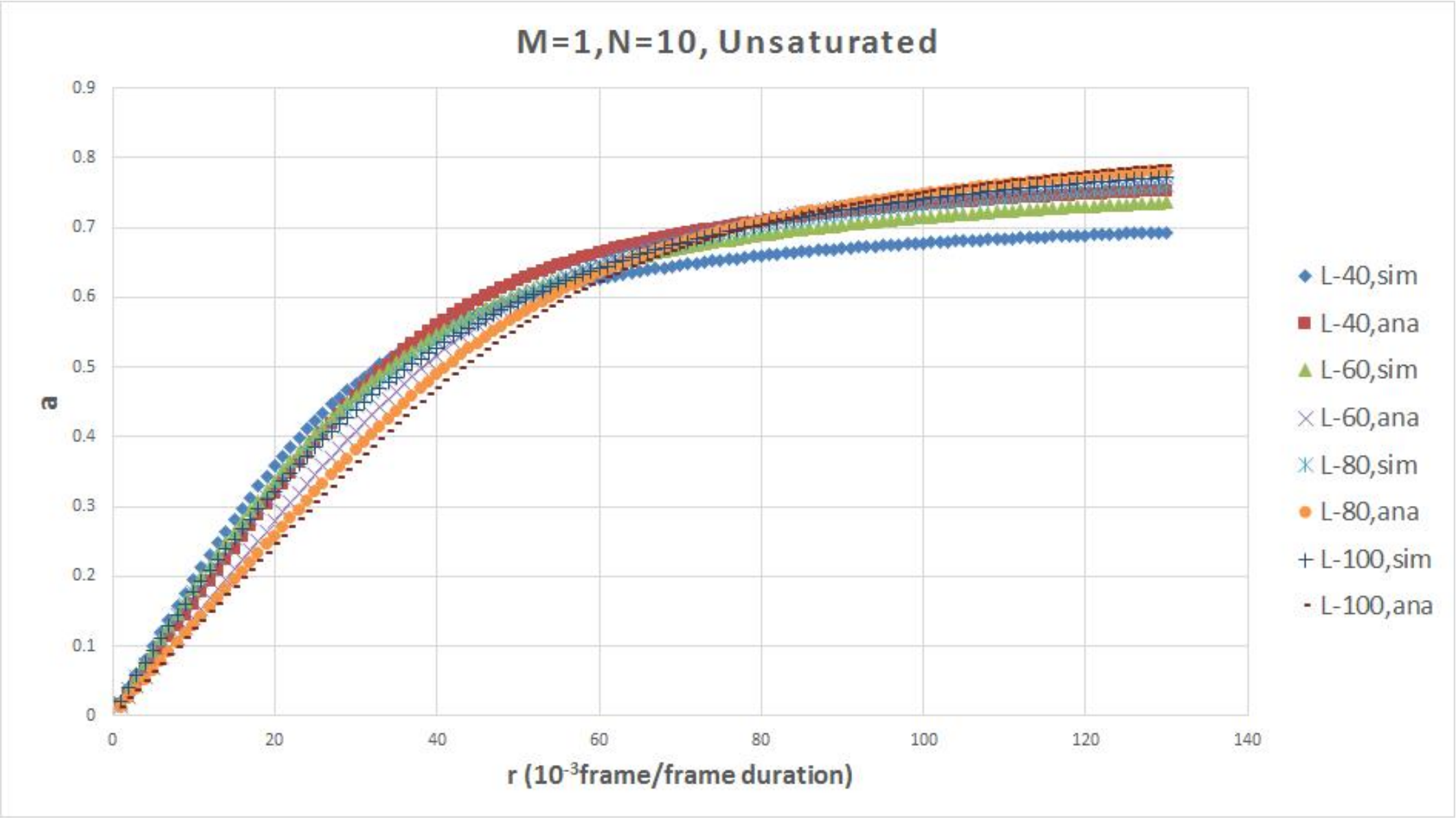

(b) Changes in $a$ for different frame lengths

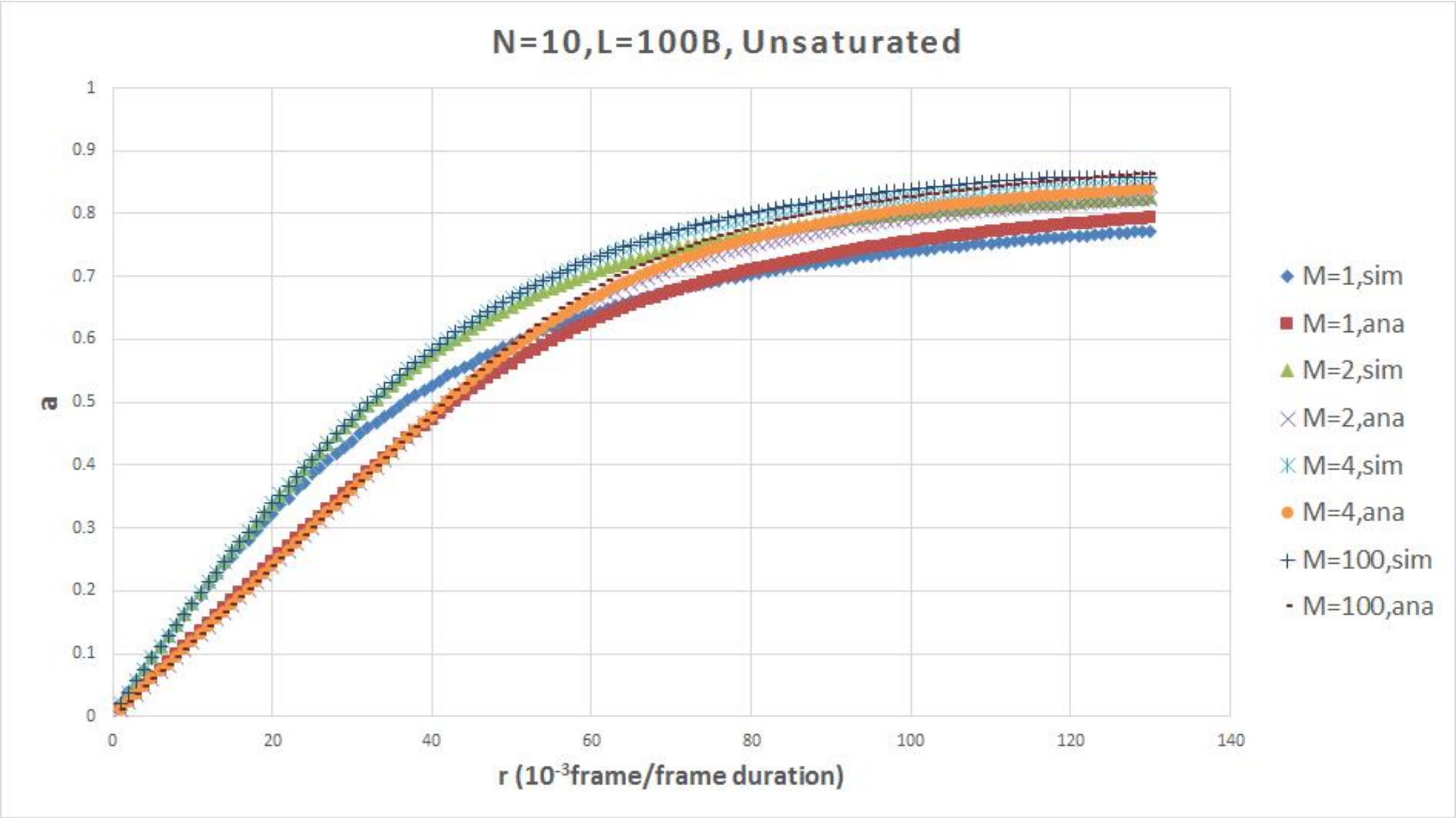

(c) Changes in $a$ for different buffer sizes

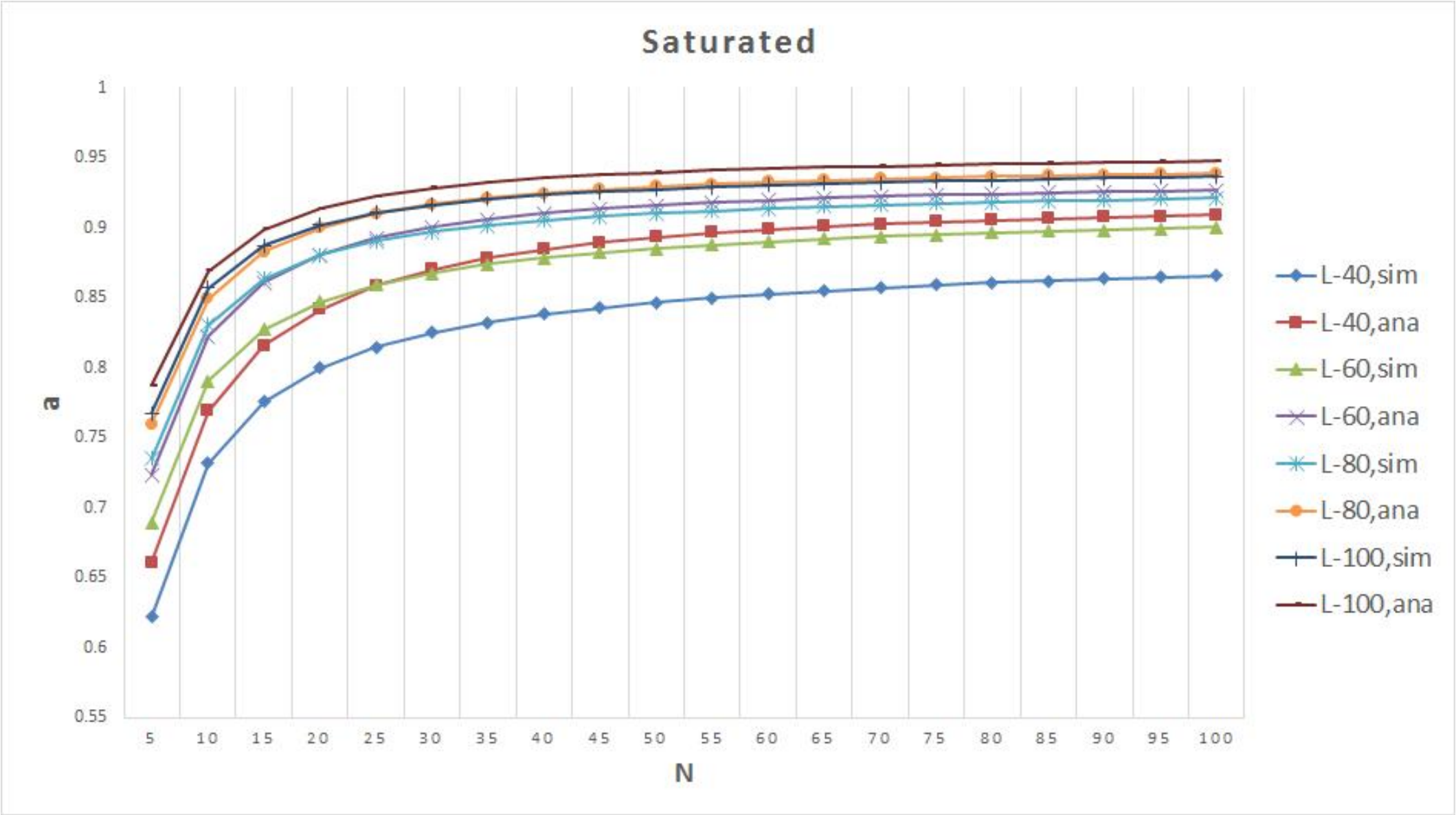

(d) Changes in $a$ for different frame lengths and saturated traffic

**Fig. 7** Probability of sensing a busy channel

## 4.2 Throughput analysis

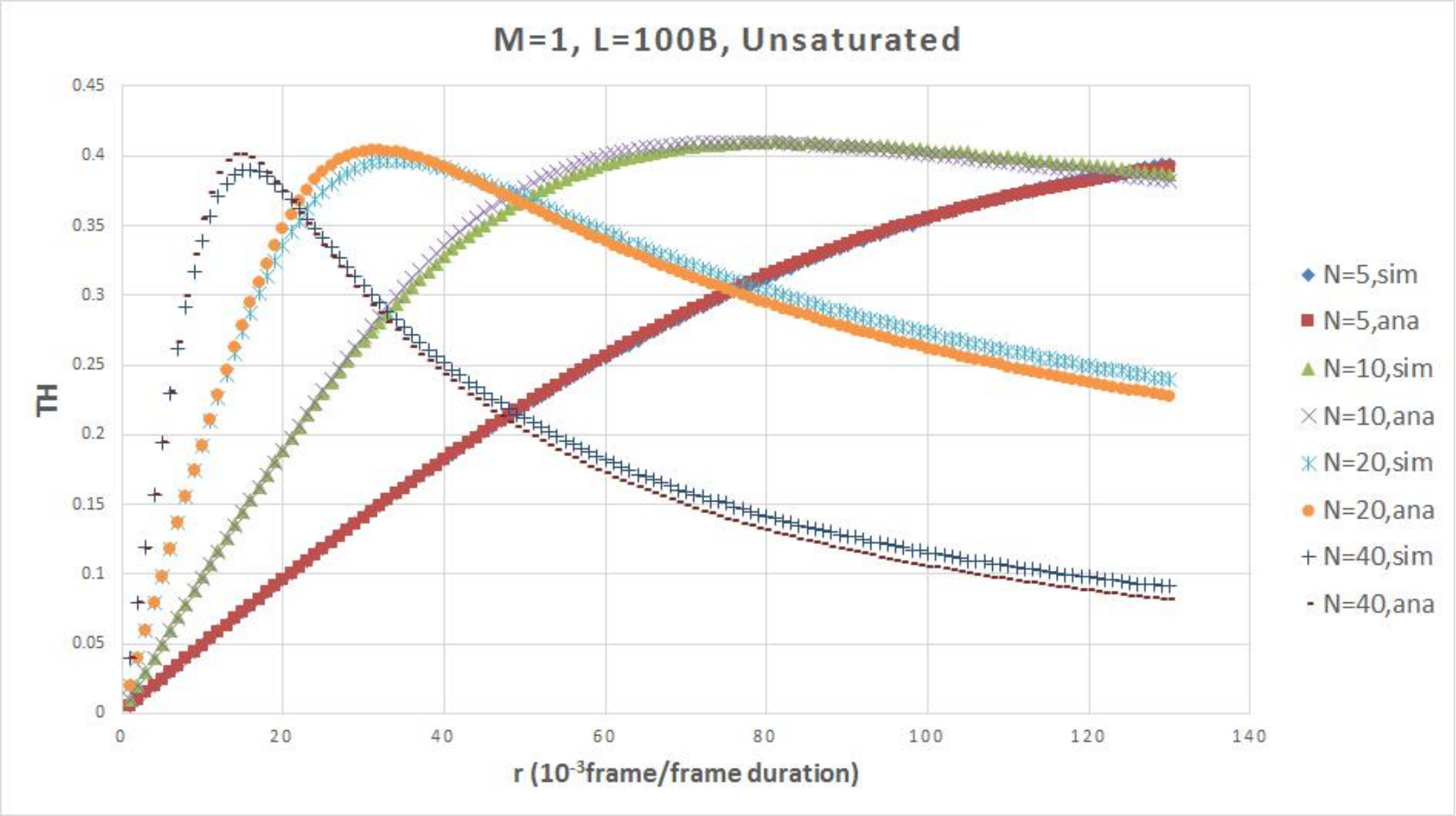

(a) Changes in TH for different node numbers

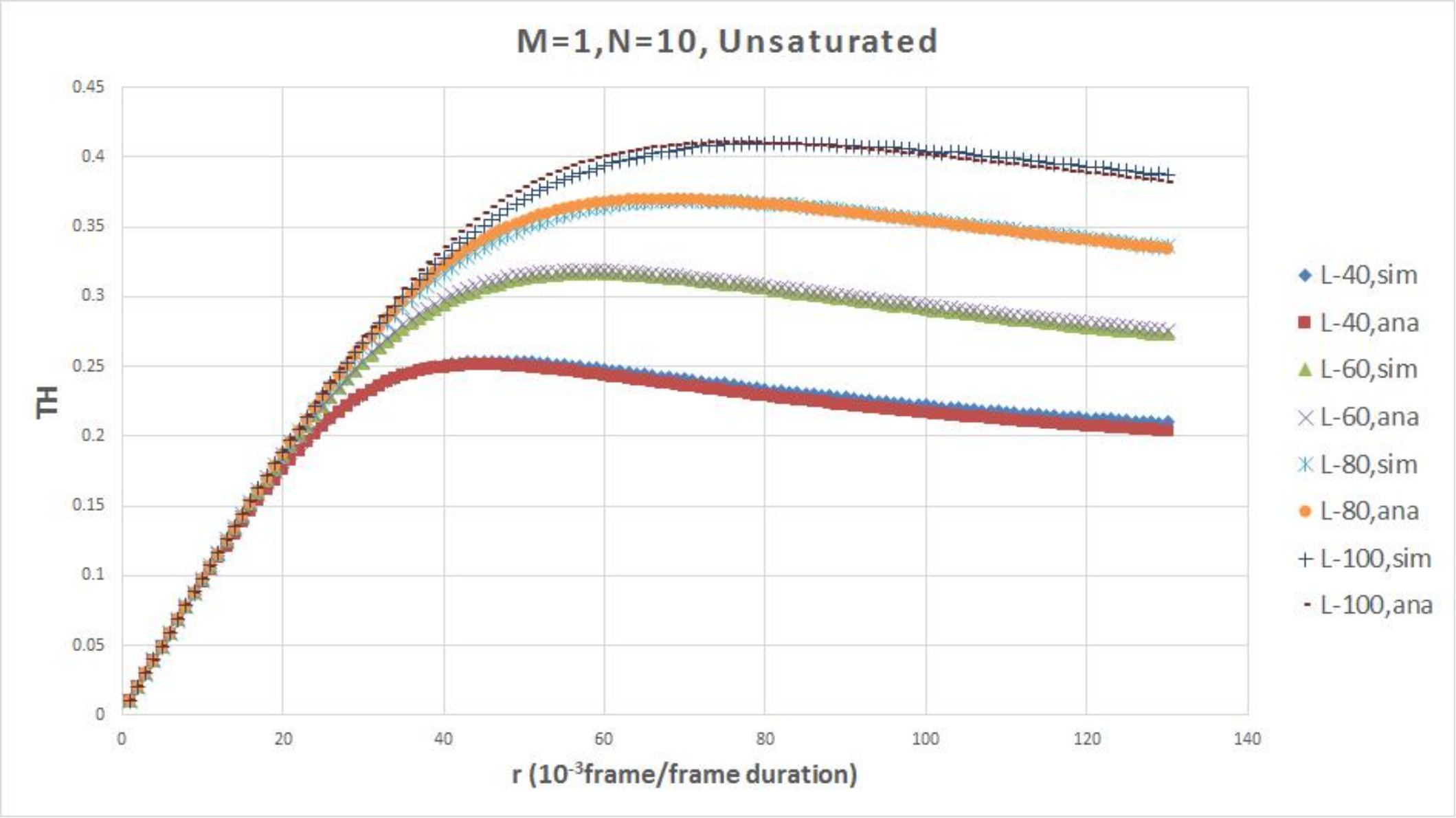

(b) Changes in TH for different frame lengths

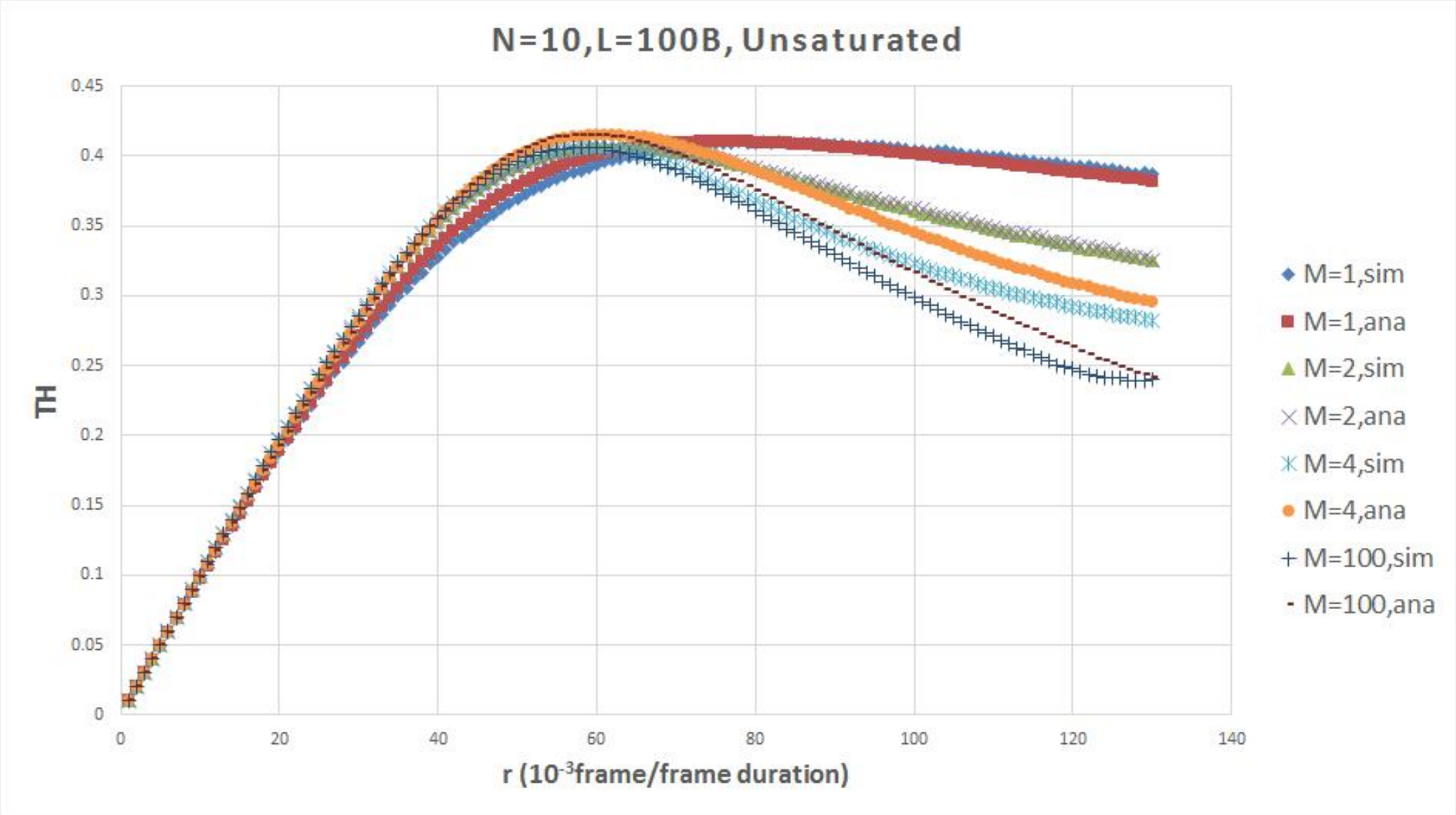

(c) Changes in TH for different buffer sizes

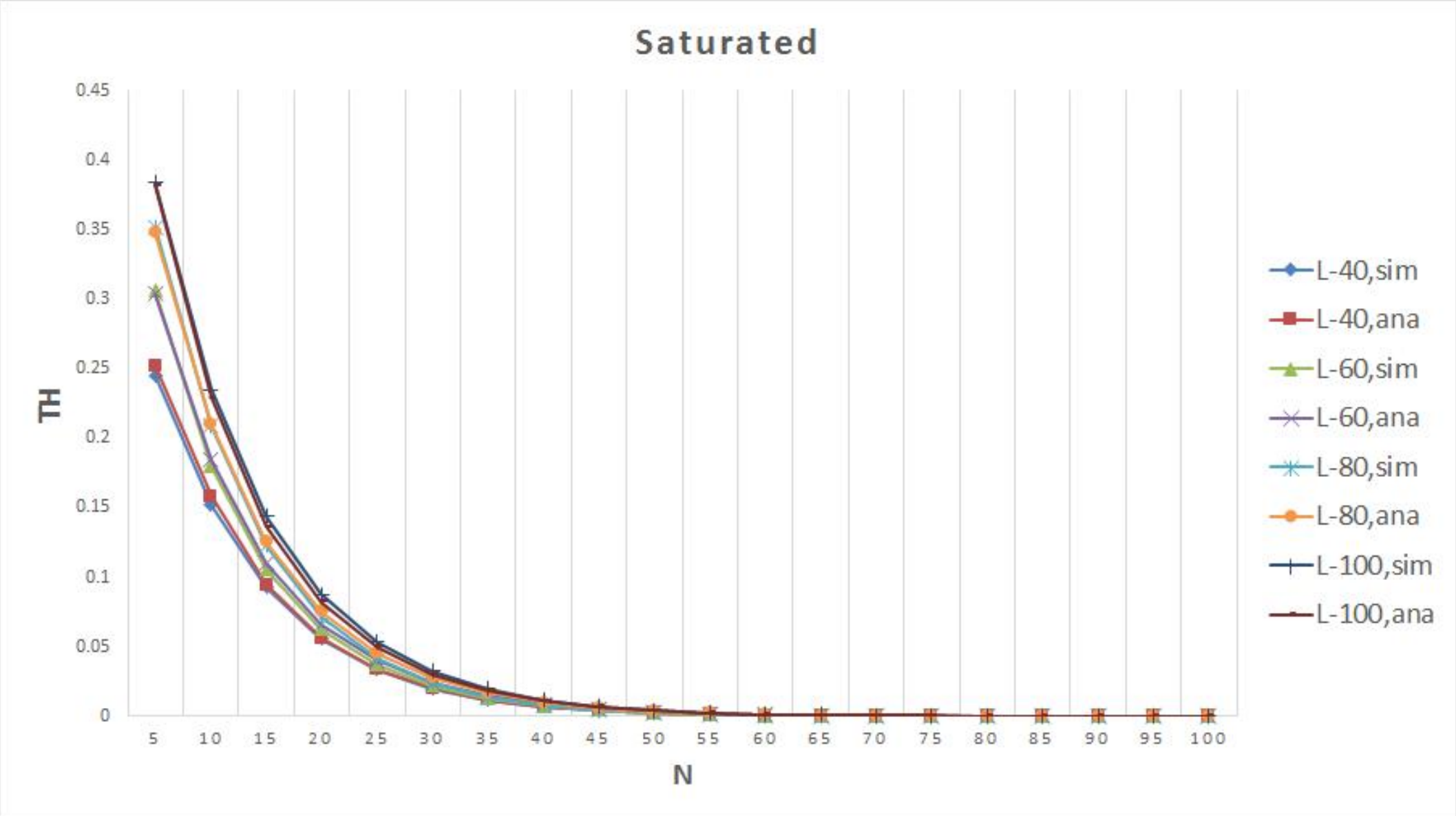

(d) Changes in TH for different frame lengths and saturated traffic

**Fig. 8** TH: the normalized throughput of the network

According to Equation (16), the throughput of the system is determined by the node number, packet length, and CCA probability. As shown in Figs. 8(a)–(c), the throughput shows a downward trend after the first rise as $r$ increases. After the throughput reaches its highest value, the packet collision probability increases because more packets are waiting to be sent, and as a result, the throughput then decreases.

The influence of node number $N$ on the probability of sensing a busy channel is illustrated in Fig. 8(a), where $L$ = 100 bytes, $r \in [0.001, 0.13]$, and M = 1. The influence of packet length $L$ on the probability of sensing a busy channel is illustrated in Fig. 8(b), where $N = 10$, $r \in [0.001, 0.13]$, and M = 1. With a fixed arrival rate, *TH* increases as L increases. The simulation results are larger than the analytical results. The influence of MAC buffer size on the CCA probability is illustrated in Fig. 8(c), where $L = 100$ bytes, $r \in [0.001, 0.13]$, and N = 10. The figure shows that the four *TH* curves are very close before reaching the maximum value, and the larger MAC buffer size leads to a *TH* reaching the maximum value more quickly. The influence of $N$ and $L$ on *TH* with saturated traffic is illustrated in Fig. 8(d). The figure shows that *TH* increases as $L$ increases.

In Fig. 8, TH increases as r, N, L, and M increase. With unsaturated traffic, TH can become saturated as r and N increase. With saturated traffic, TH decreases sharply as N increases. With unsaturated traffic, compared with Figs. 6–7, TH and $a$ follow similar trends to reach the maximum value, but $\tau$ still does not reach its maximum value. This indicates that the node still does not reach a busy state even though the channel has been saturated. Correspondingly, for a given node number, the packet sending rate should not continue to increase when approaching the maximum throughput.

## 4.3 Reliability analysis

According to Equation (29), the probability of a successful transmission of a data frame is determined by the node number, probability of sensing a busy channel, and CCA probability. The influence of node number $N$ on *PS* is illustrated in Fig. 9(a), where $L = 100$ bytes, $r \in [0.001, 0.13]$, and M = 1. *PS* decreases as $N$ and $r$ increase. The influence of packet length $L$ on the probability of sensing a busy channel is illustrated in Fig. 9(b), where $N = 10$, $r \in [0.001, 0.13]$, and M = 1. With a fixed arrival rate, *PS* increases as $L$ increases. The influence of MAC buffer size on *PS* is illustrated in Fig. 9(c), where $L = 100$ bytes, $r \in [0.001, 0.13]$, and N = 10. *PS* decreases as the MAC buffer size increases at a fixed packet arrival rate. The difference of simulation results and analytical results obtain larger as the MAC buffer size gets larger. The influence of $N$ and $L$ on *PS* with saturated traffic is illustrated in Fig. 9(d). The figure shows that *PS* decreases as L increases.

In Fig. 9, because the simulation result of $a$ is larger than the analytical result of $a$, the simulation result of PS is smaller than the analytical result of PS.

In Fig. 9, PS increases as r, N, and L increase, whereas it decreases as M increases. With unsaturated traffic, it is hard to keep a high PS when N > 10, even with a low rate of transmission. Therefore, in applications that require a higher delivery ratio, it is essential to keep a low sending rate or a few nodes. With saturated traffic, PS is small when N > 5.

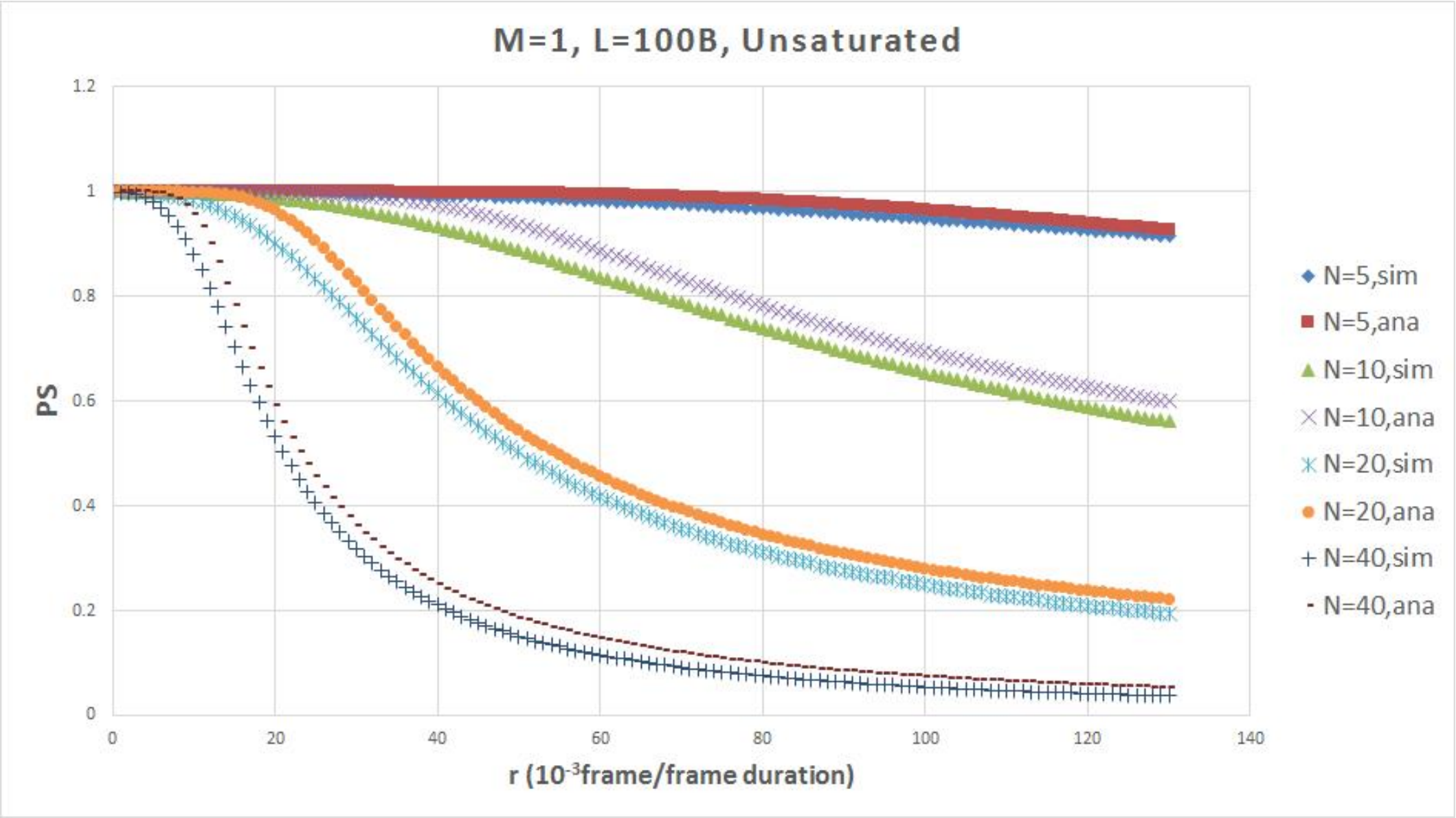

(a) Changes in PS for different node numbers

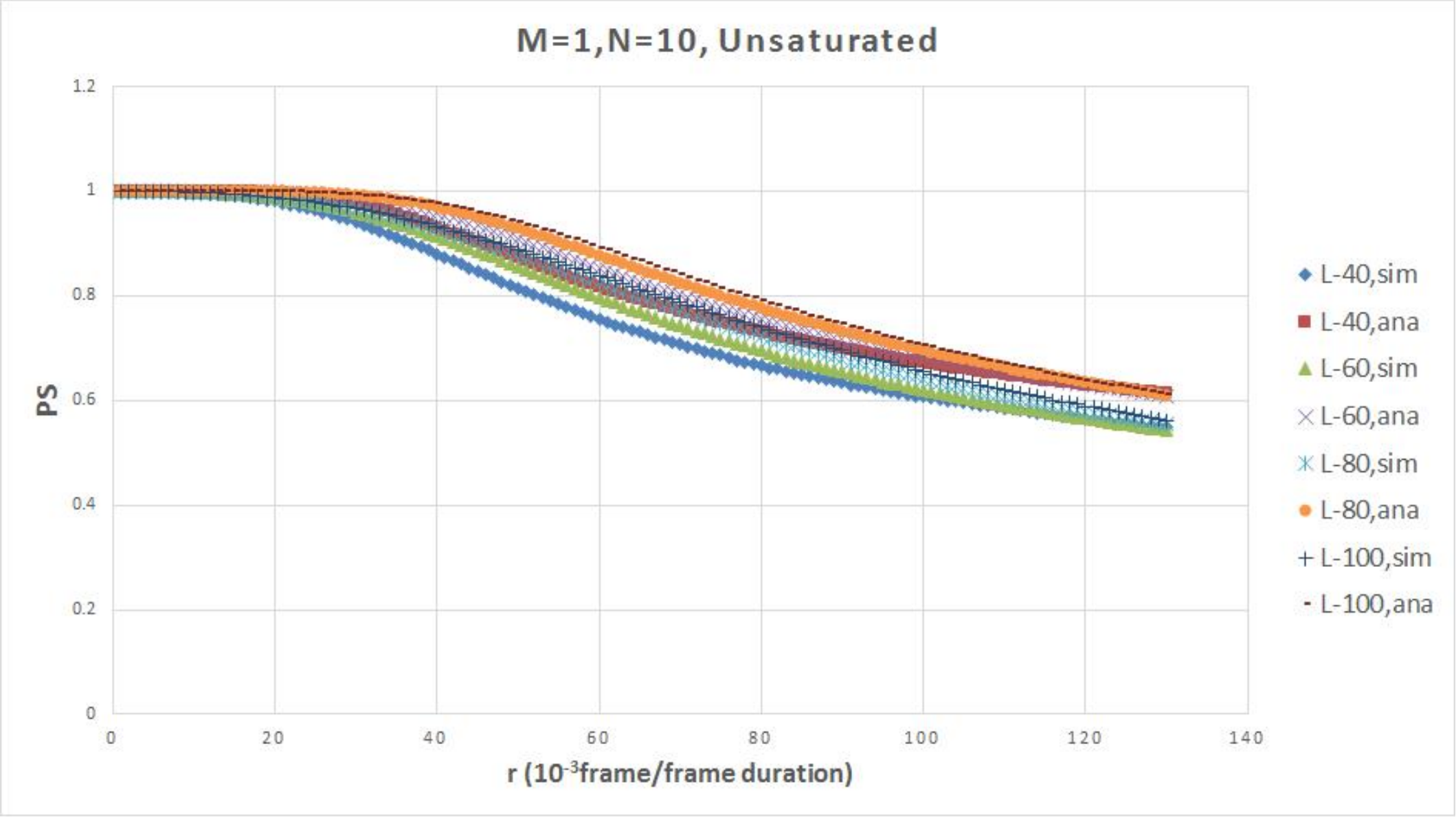

(b) Changes in PS for different frame length

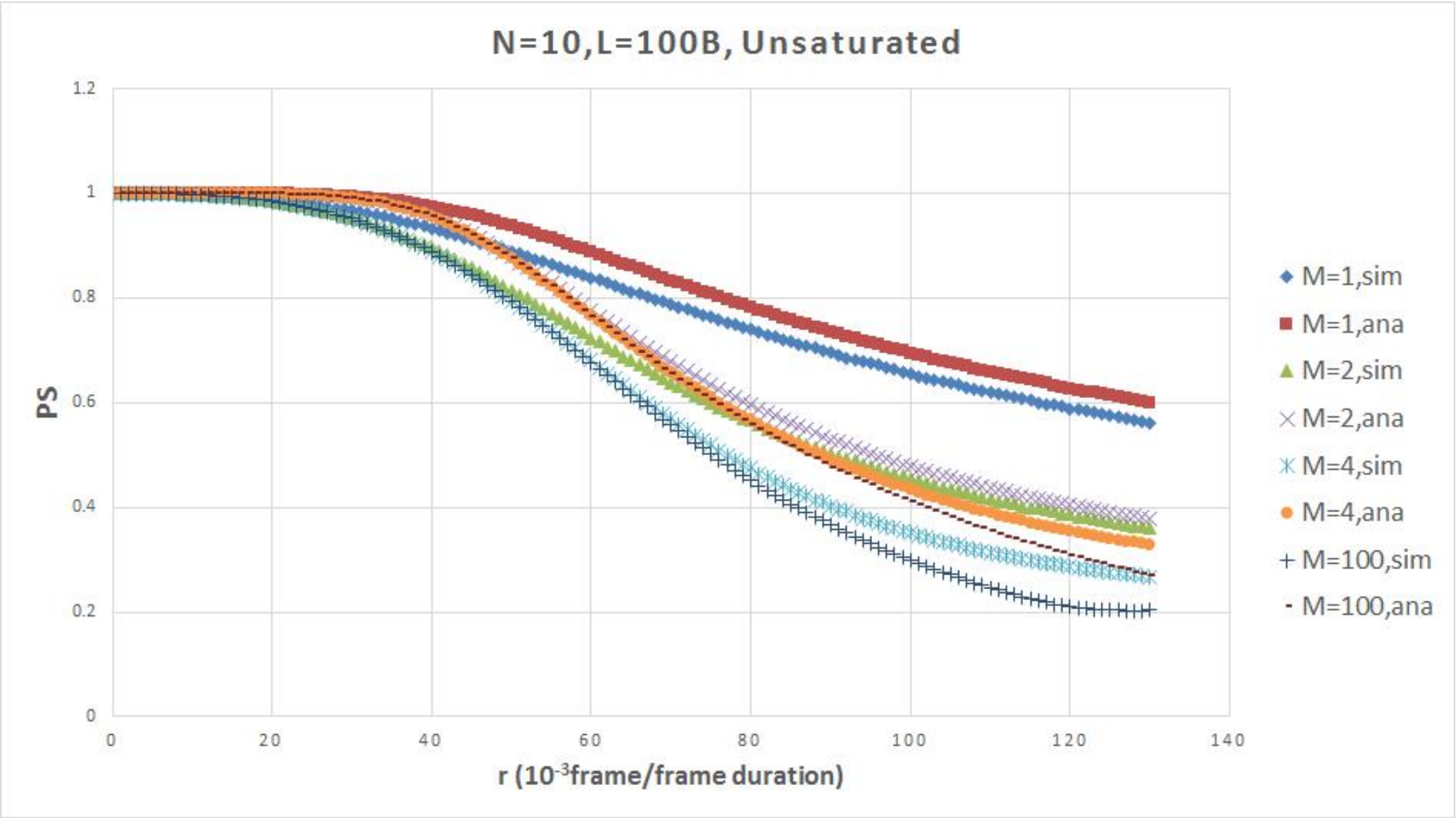

(c) Changes in PS for different buffer size

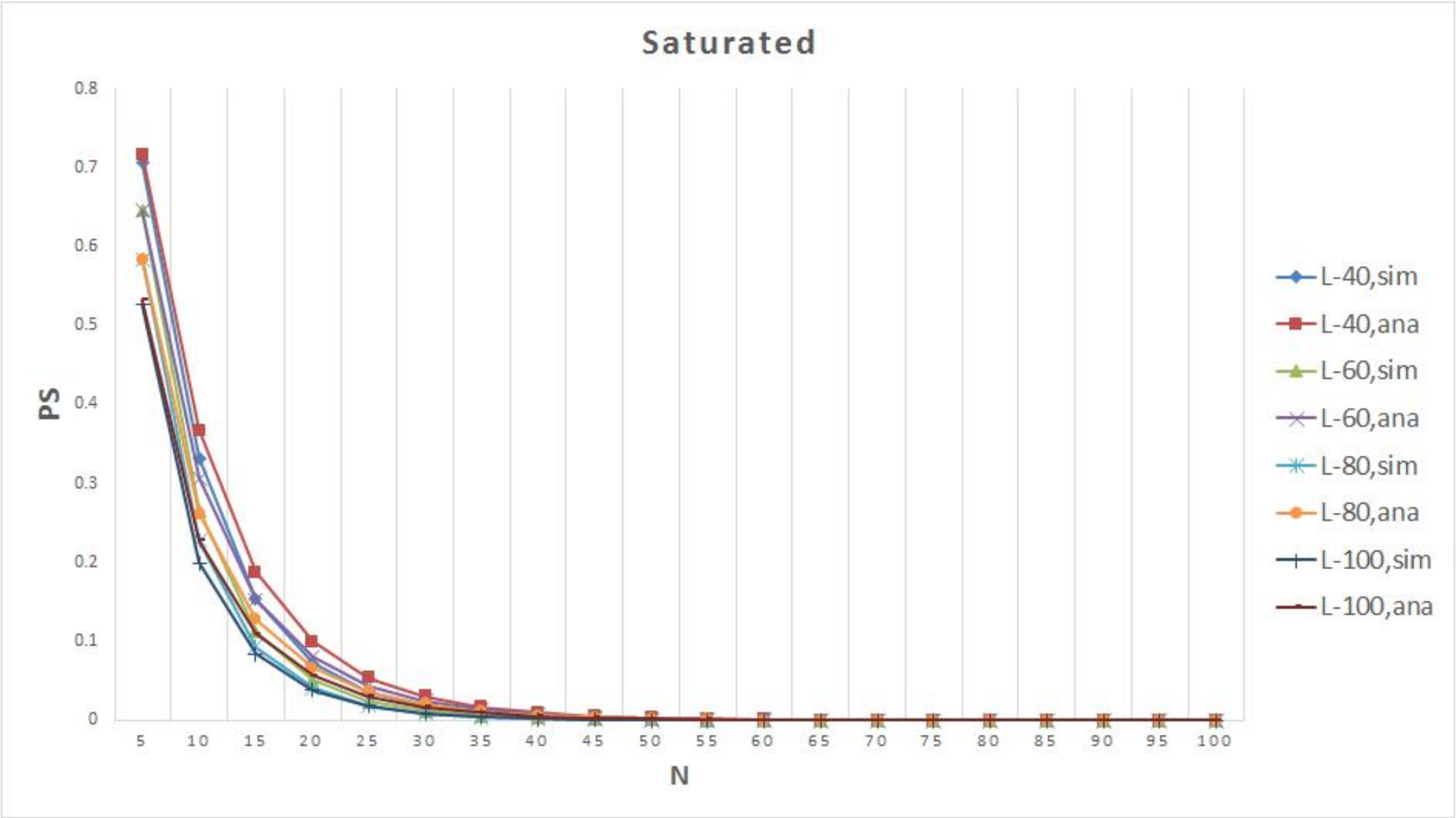

(d) Changes in PS for different frame length in saturated case

**Fig. 9** PS: the probability of successful data transmission

# 4.4 Delay analysis

According to Equations (30–31) and (35–36), the delay of data frame transmission is determined by the node number, packet length, MAC buffer size, probability of sensing a busy channel, and CCA probability.

The influence of node number $N$ on $TS$ is illustrated in Fig. 10(a), where $L$ = 100 bytes, $r \in [0.001,0.13]$, and M = 1. $TS$ increases as $N$ and $r$ increase. As N increases, there are more nodes sensing the channel, leading to the increase of CCA probability failure or packet collision and, correspondingly, the total delay of successful data frame transmissions will be increased. The influence of packet length *L* on *TS* is illustrated in Fig. 10(b), where $N = 10$, $r \in [0.001, 0.13]$, and M = 1. At a fixed arrival rate, *TS* increases as L increases. The influence of MAC buffer size on TS is illustrated in Fig. 10(c), where $L = 100$ bytes, $r \in [0.001, 0.13]$, and N = 10. *TS* increases as the MAC buffer increases at a fixed packet arrival rate. The influence of *N* and *L* on *TS* with saturated traffic is illustrated in Fig. 10(d). The figure shows that *TS* increases as L increases.

In Fig. 10, the simulation result of TS is larger than the analytical result. For the same the reason that the analytical result of PS is larger than the simulation result in Fig. 9, it needs less time for a successful data packet transmission.

In Fig. 10, TS increases as r, N, and M increase, whereas it decreases as L increases. When TS reaches the maximum value, TS remains steady. Because the maximum retransmission number is three, successful transmissions will go through three collisions at most. These results show that more time is spent for the successful transmissions of data frames as r, N, and M increase.

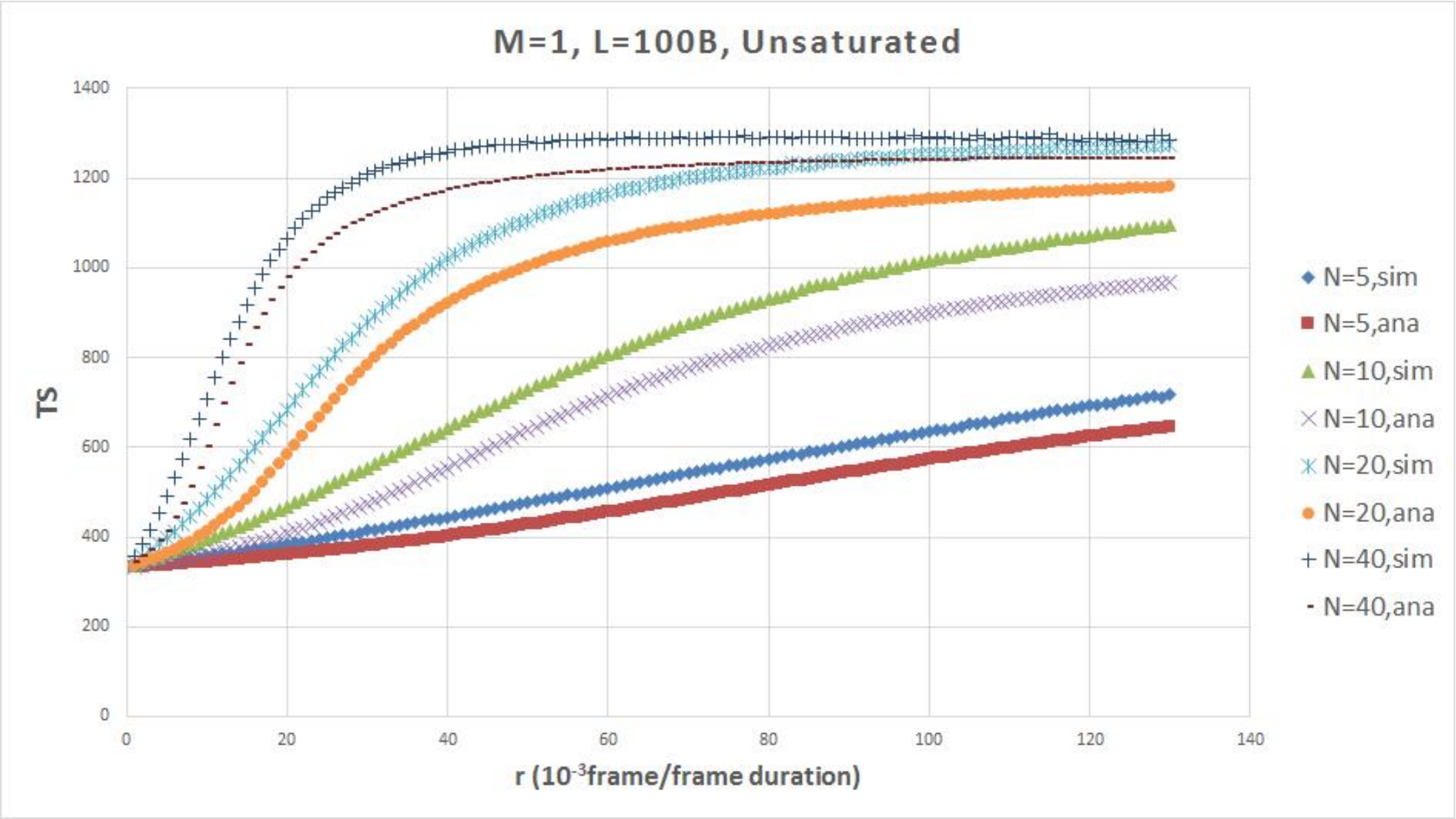

(a) Changes in TS for different node numbers

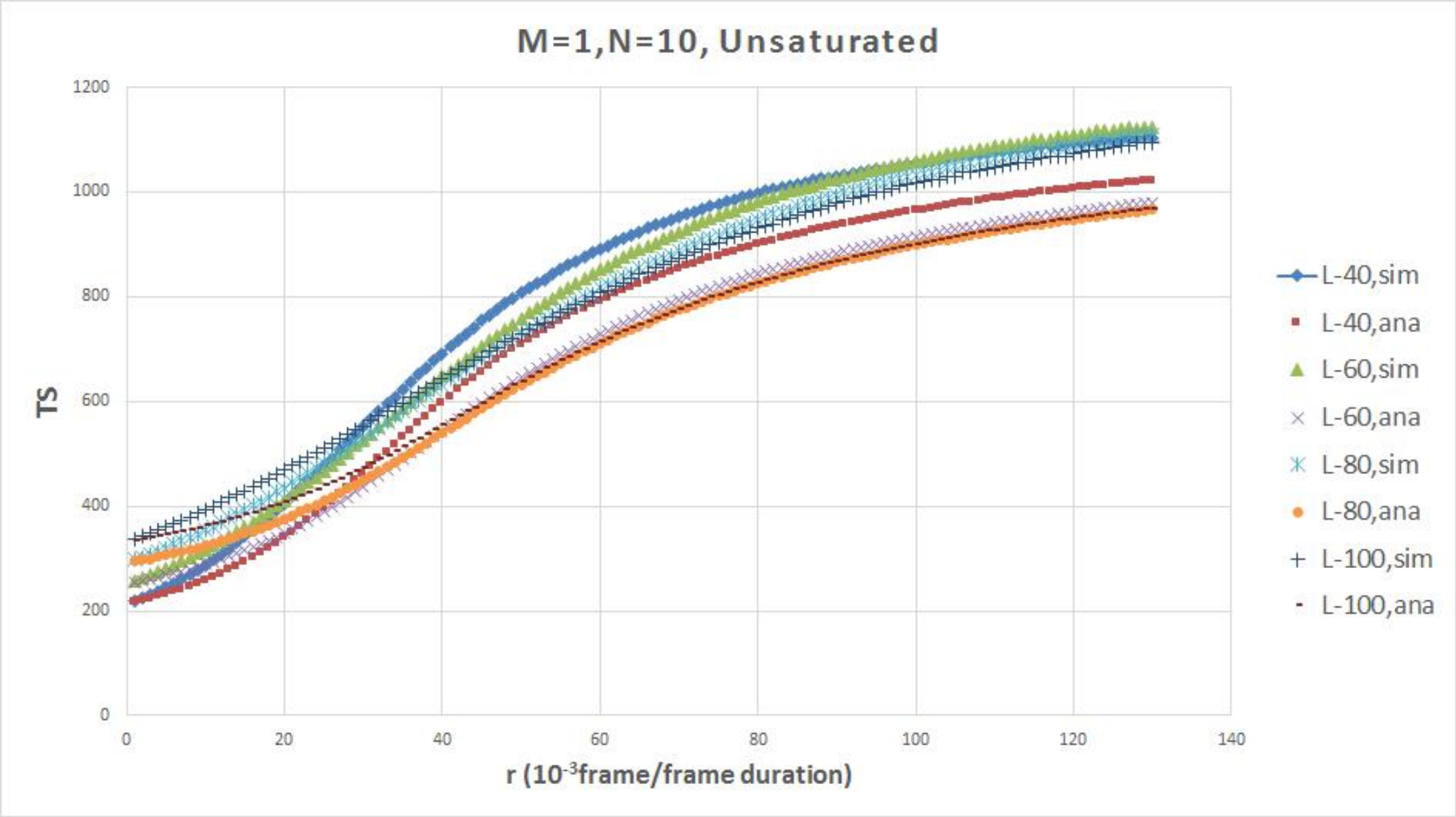

(b) Changes in TS for different frame lengths

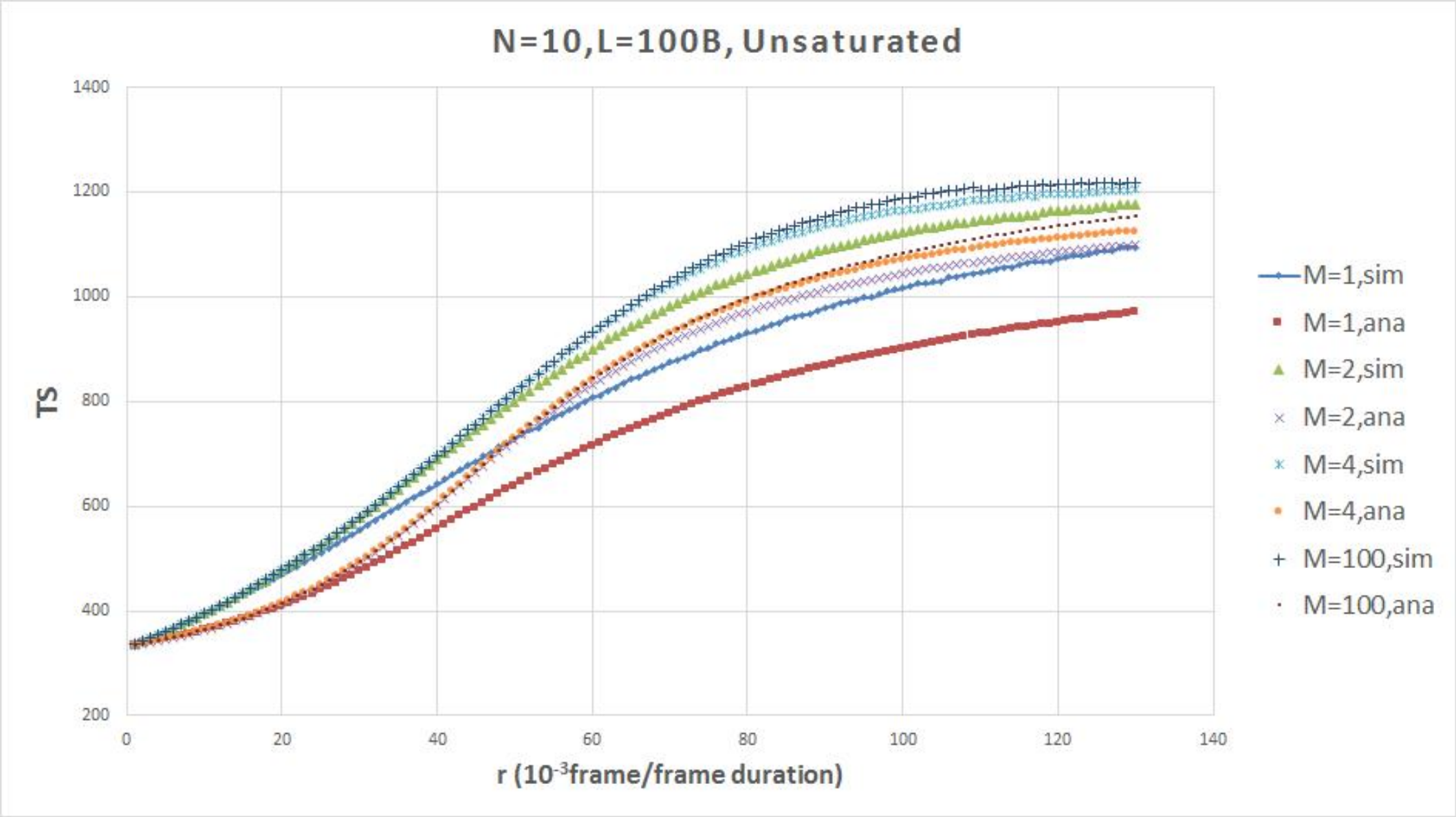

(c) Changes in TS for different buffer sizes

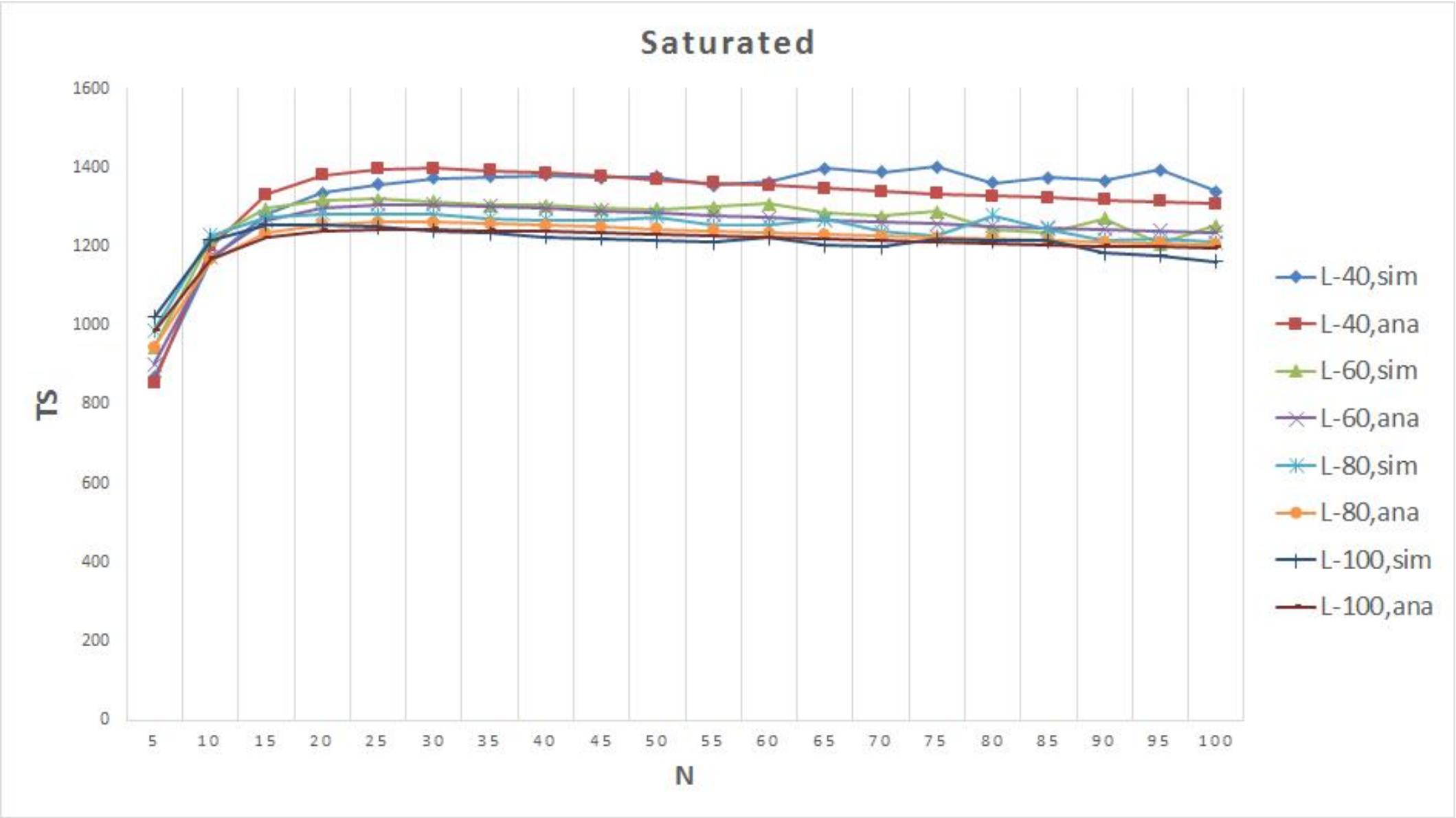

(d) Changes in TS for different frame lengths in the saturated case

**Fig. 10** TS: the average time of successful data transmissions without queueing delay

The influence of MAC buffer size on *TSW* is illustrated in Fig. 11, where $L$ = 100 bytes, $r \in$ [0.001,0.13], and N = 10. For a fixed packet arrival rate, *TSW* increases as the MAC buffer increases. The results of Fig. 11 show that an increase in MAC buffer size will increase the queueing delay of the packets, which can be unacceptable in a time-constrained application.

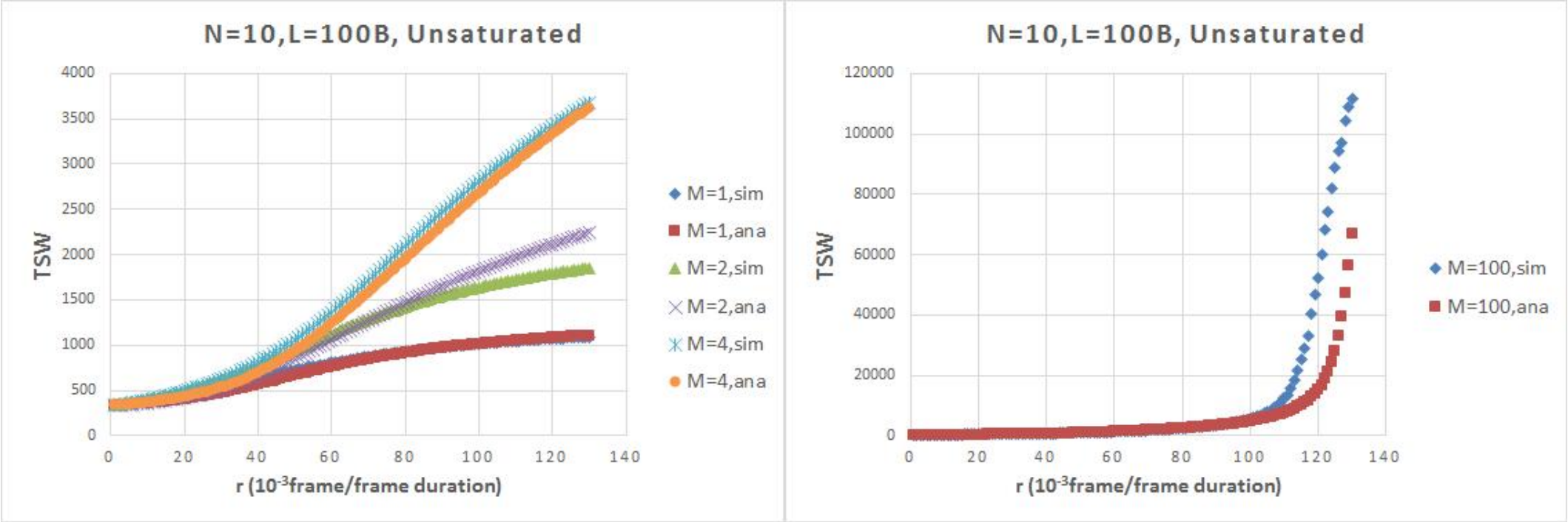

**Fig. 11** TSW: the average time of successful data transmission with queueing delay

The influence of node number $N$ on *TVS* is illustrated in Fig. 12(a), where $L$ = 100 bytes, $r \in$ [0.001,0.13], and M = 1. *TVS* increases as N and $r$ increase. The influence of packet length $L$ on *TVS* is illustrated in Fig. 12(b), where $N$ = 10, $r \in$ [0.001,0.13], and M = 1. For a fixed arrival rate, *TVS* increases as L increases. The simulation results are larger than the analytical results. The influence of MAC buffer size on *TVS* is illustrated in Fig. 12(c), where $L$ = 100 bytes, $r \in$ [0.001,0.13], and N = 10.

At a fixed packet arrival rate, *TVS* increases as the MAC buffer size increases. The influence of *N* and *L* on *TVS* with saturated traffic is illustrated in Fig. 12(d). The figure shows that *TVS* increases as L increases.

In Fig. 12, the simulation results are larger than the analytical results, because the simulation results of PS are smaller than its analytical results, and the data transmission delay is increased by the increase of CCA failure or retransmission. In Fig. 12, *TVS* increases as r, N, and M increase, whereas it decreases with increases in L. Compared to Fig. 10, TVS is larger than TS. This is because unsuccessful data transmissions take more time than successful data transmissions.

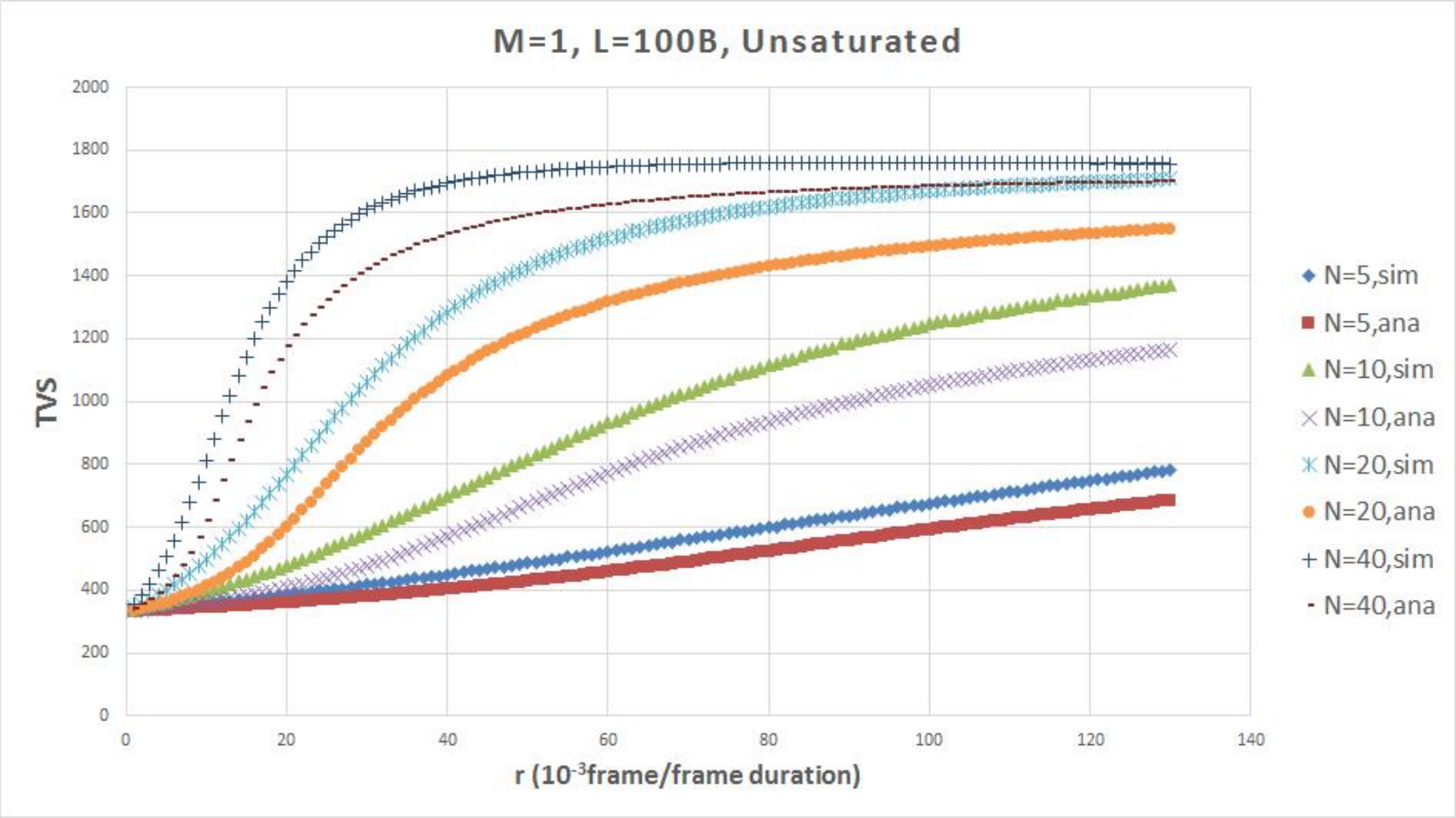

(a) Changes in TVS for different node numbers

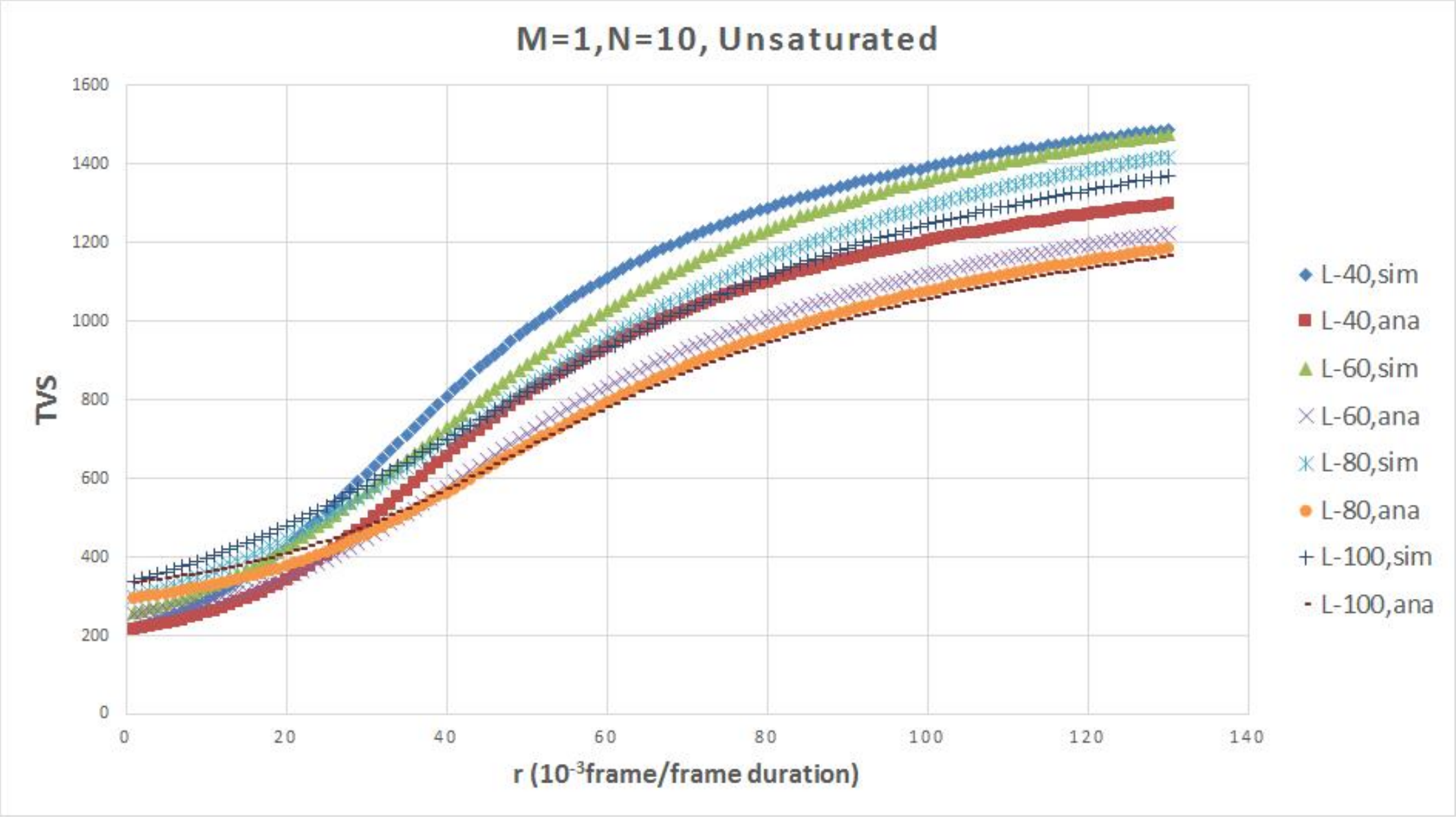

(b) Changes in TVS for different frame lengths

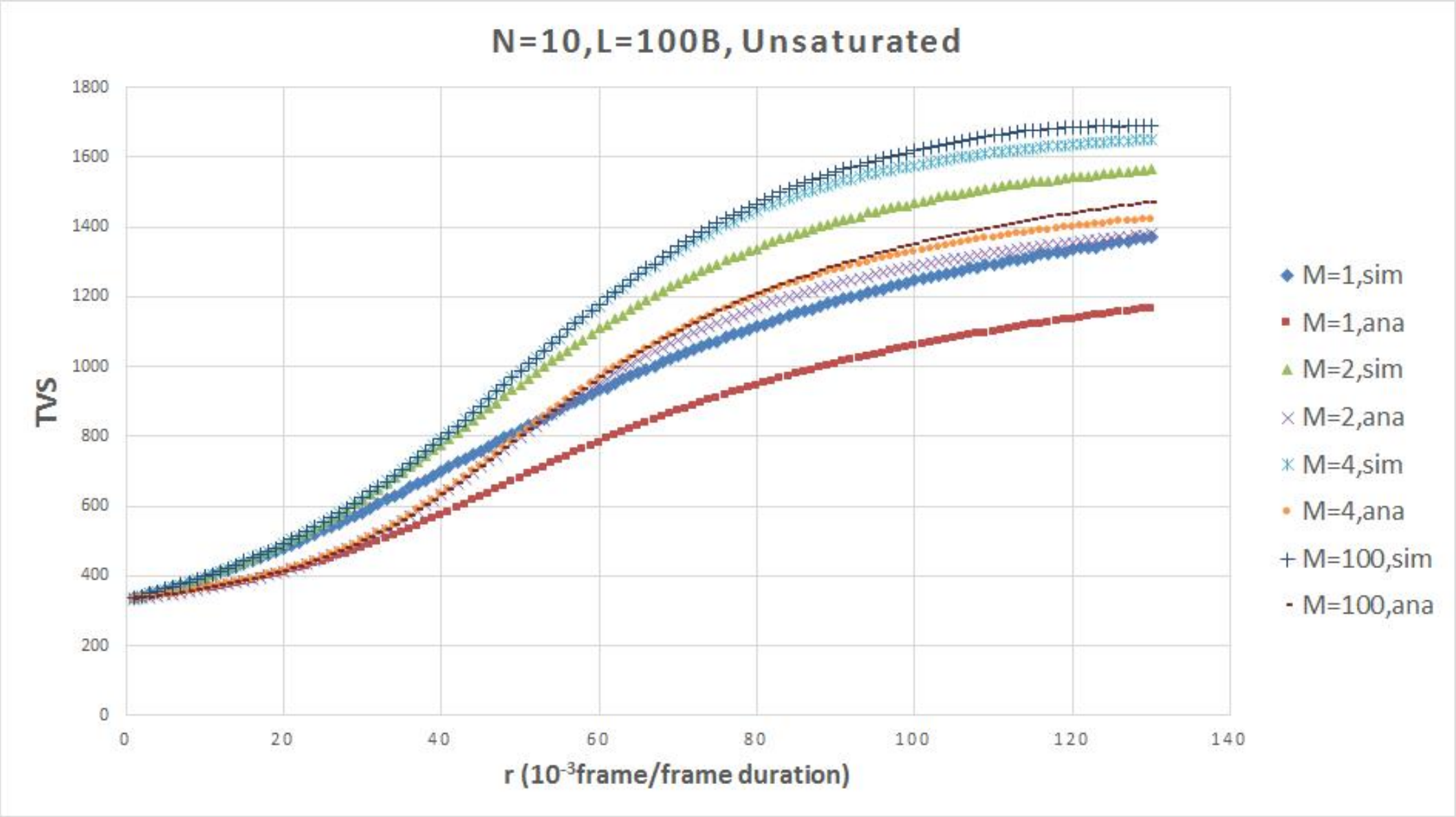

(c) Changes in TVS for different buffer sizes

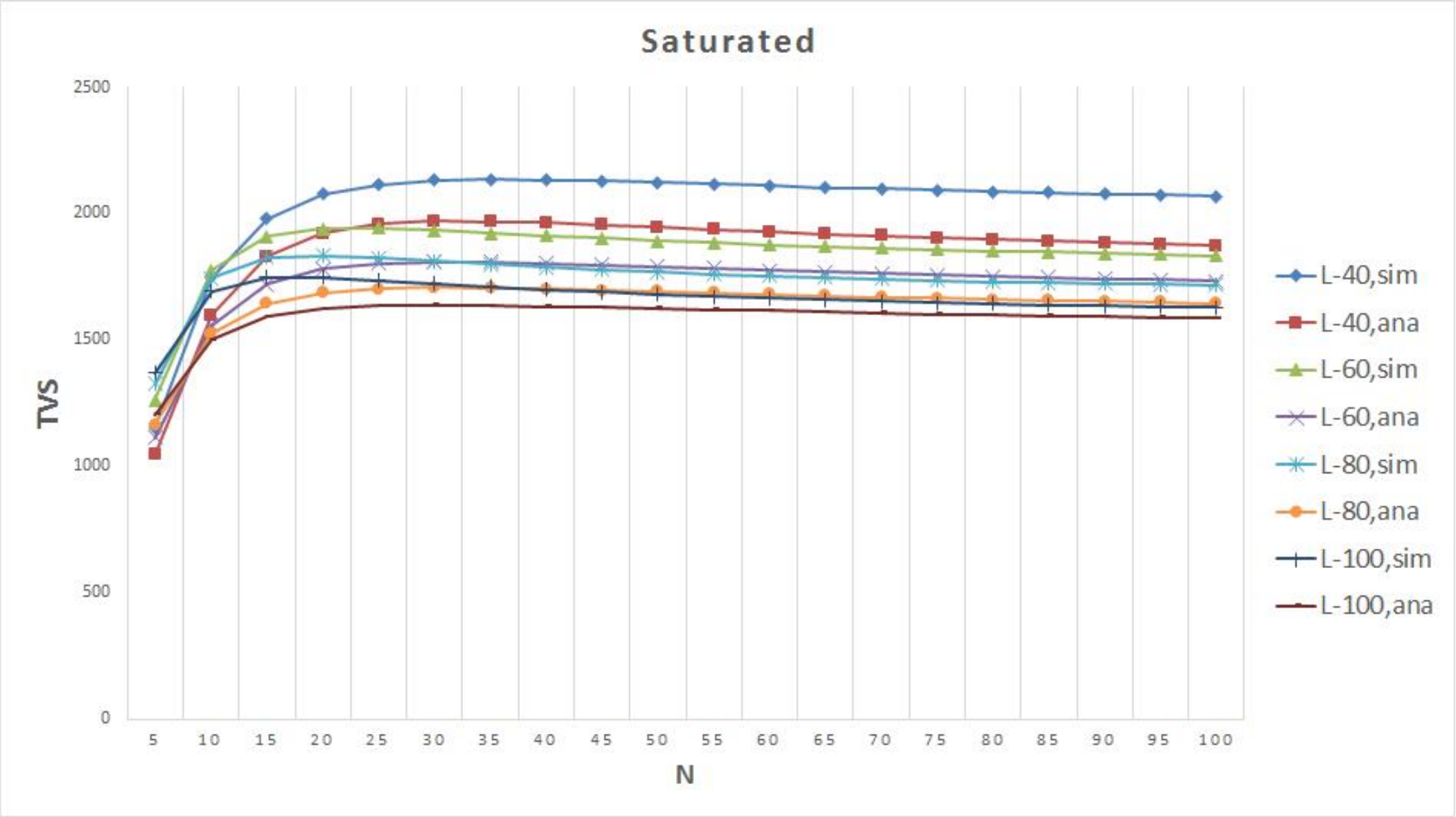

(d) Changes in TVS for different frame lengths for the saturated case

**Fig. 12** TVS: the average time of data transmission without queueing delay

The influence of the MAC buffer size on *TVSW* is illustrated in Fig. 13, where $L$ = 100 bytes, $r \in [0.001,0.13]$, and N = 10. At a fixed packet arrival rate, *TVSW* increases as the MAC buffer size increases. The results of Fig. 13 show that the MAC buffer size increases increase the queueing delay of the packets, which may be unacceptable in time-constrained applications.

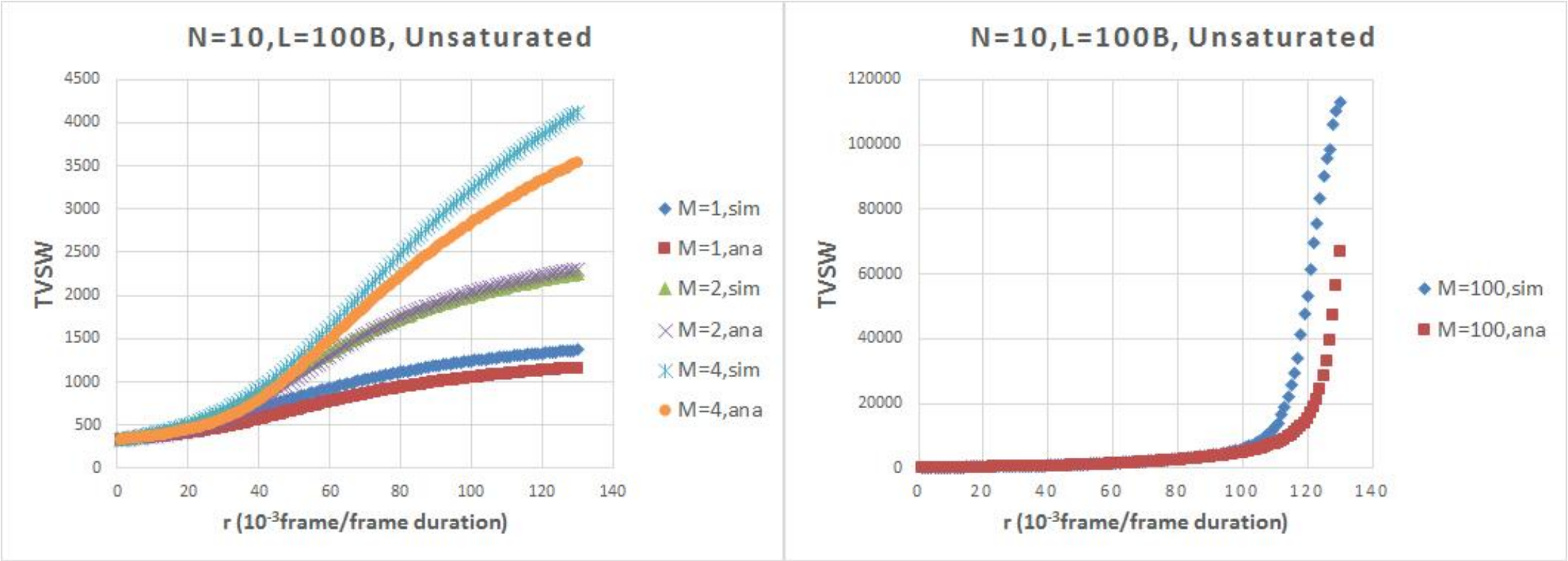

**Fig. 13** TVSW: the average time of data transmission with queueing delay

## 4.5 Discussion of the Markov chain model

We propose three Markov chain models under different conditions that are based on a fixed channel model. Eight performance parameters are derived from these models. For *TH*, the analytical results are in close accordance with the simulation results. However, there exists a small difference between the analytical results and the simulation results for $\tau$, *a, PS, TS, TVS,* and *TVSW.*

The influence of *r, L, M,* and *N* on these performance parameters are shown in Figs. 6–12. Based on the analytical model, we can derive the expressions of the other parameters using $\tau$ and $a$. For this reason, the analytical results of the other parameters are dependent on $\tau$ and $a$. Based on this conclusion, we use the simulation results of $\tau$ and $a$ to obtain the other parameters, and hence the analytical results of the other parameters of *TS* and *TVS* will get closer to the simulation results.

Here, we think the difference between the simulation and analytical results has two causes. The first cause is the imperfection of the overall model. For simplification, the channel models described in Section 3.2 assume that there are no nodes sensing a channel in the *wack* state, as shown in Fig. 4. In fact, there are very few nodes sensing the channel immediately after packet collision. This is the reason why the analytical results of $\tau$ are a little smaller than the simulation results and the analytical results of $a$ are smaller than the simulation results. Thus, the analytical results of *PS* are a little higher than the simulation results. Further, *TS, TVS, TSW,* and *TVSW* are consequently a little smaller than the simulation results.

Although we attempted to put the *wack* state into the *IDLE1* status in Fig. 4, the difference between the simulation and analytical results grew. This is because there are at most $N-2$ nodes sensing the channel in the *wack* state. In a non-beacon network, the channel model may not always be in a steady state, so it is an approximation.

The second cause is that the basic time unit in our models is 1 symbol (16 μs), and all collisions are considered. Hence, these considerations add to the complexity of the models, which leads to analytical expressions that are only approximately close to the simulation results, even if the simulation of $\tau$ and $a$ are used to derive other parameters.

With unsaturated traffic, the MAC buffer size increases the average delay of packet transmission, and a large MAC buffer size does not improve the *PS* and *TH*. Increasing *L* can improve the *PS* and *TH* and reduce the average transmission delay. The increase of *N* and *r* raises the number of packets to send and quickly saturates the network. After the network throughput reaches its highest value, the performance of the network reduces sharply.

## 4.6 Prediction results

Based on the description of Section 3.4, we trained the BP-ANN models using the data from the analytical results of the Markov chain model with unsaturated traffic when $M$ = 1. Table 3 illustrates the results of the BP-ANN model by *R* (the coefficient of correlation) and *MSE* (the Mean Squared Error).

Table 3 Performance of the BP-ANN models

| Model name | Input vector | Output | Testing set | Number of Iterations | Time | R | MSE |
|---|---|---|---|---|---|---|---|
| **Model1** | [r,L,PS,TVS] | N | 100 | 400,000 | 48h | 0.99991 | 0.01 |
| **Model2** | [r,L,N,TVS] | PS | 100 | 440,000 | 50h | 0.99993 | $2*10^{-5}$ |
| **Model3** | [r,L,PS,N] | TVS | 100 | 460,000 | 60h | 0.99984 | $3*10^{-5}$ |

We used the simulation results to validate the results of the BP-ANN models. The comparisons are illustrated in Fig. 14.

In Fig. 14(a), the x-axis represents the number of nodes and the y-axis represents the number of nodes predicted by the BP-ANN. There are six curves in Fig. 14(a) for different conditions. The predicted *N* is larger than the simulation results of *N*. In Fig. 14(b), the x-axis represents the number of nodes and the y-axis represents the probability of successful transmission of data frame *PS*. There are 12 curves comparing the prediction results and simulation results. The predicted *PS* results are close to the simulation *PS* results. In Fig. 14(c), the x-axis represents the number of nodes and the y-axis represents *TVS*. There are 12 curves comparing the prediction results and simulation results. The predicted *N* is smaller than the simulation result of *N*.

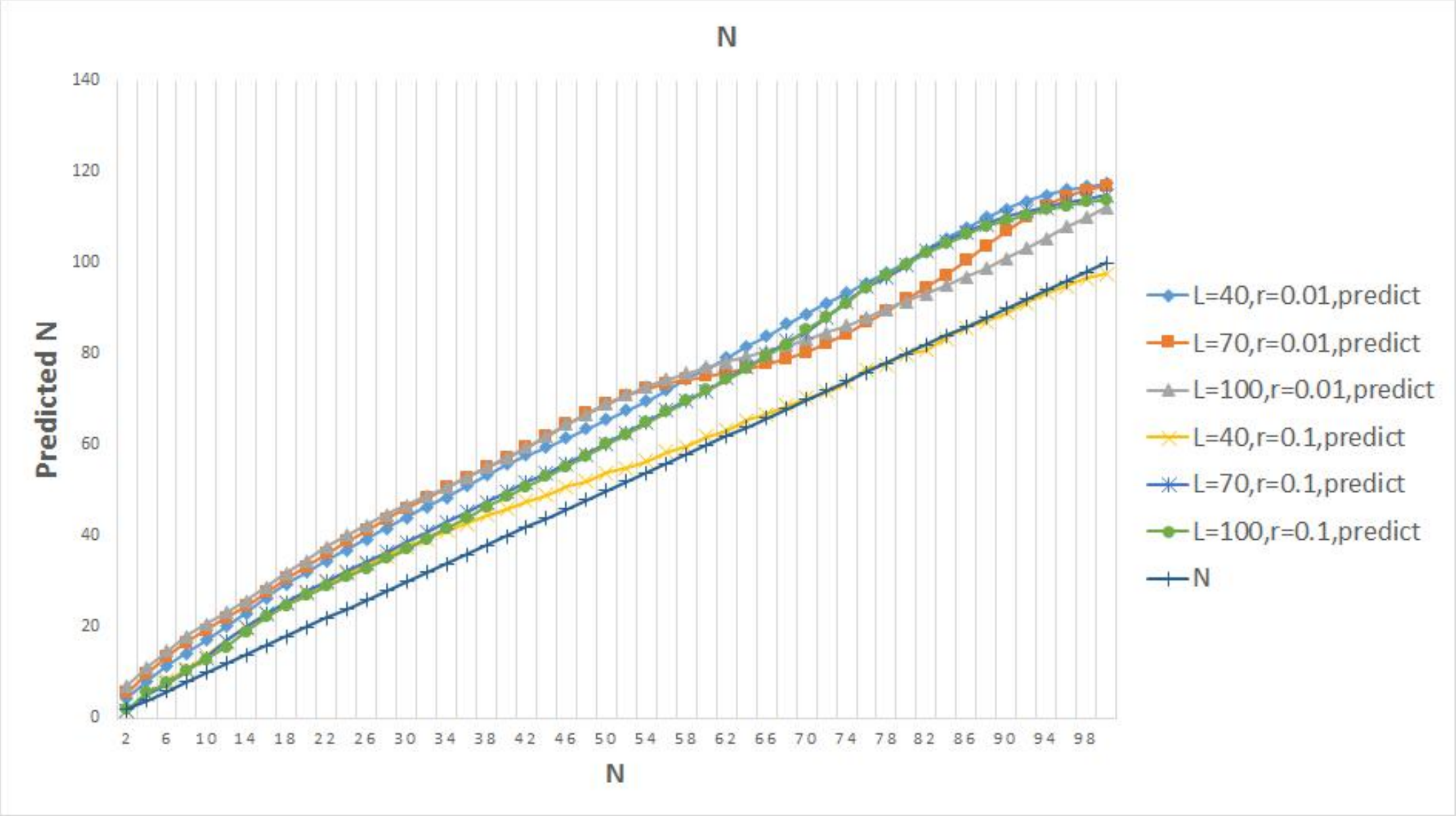

(a) Comparison of simulated and predicted *N*

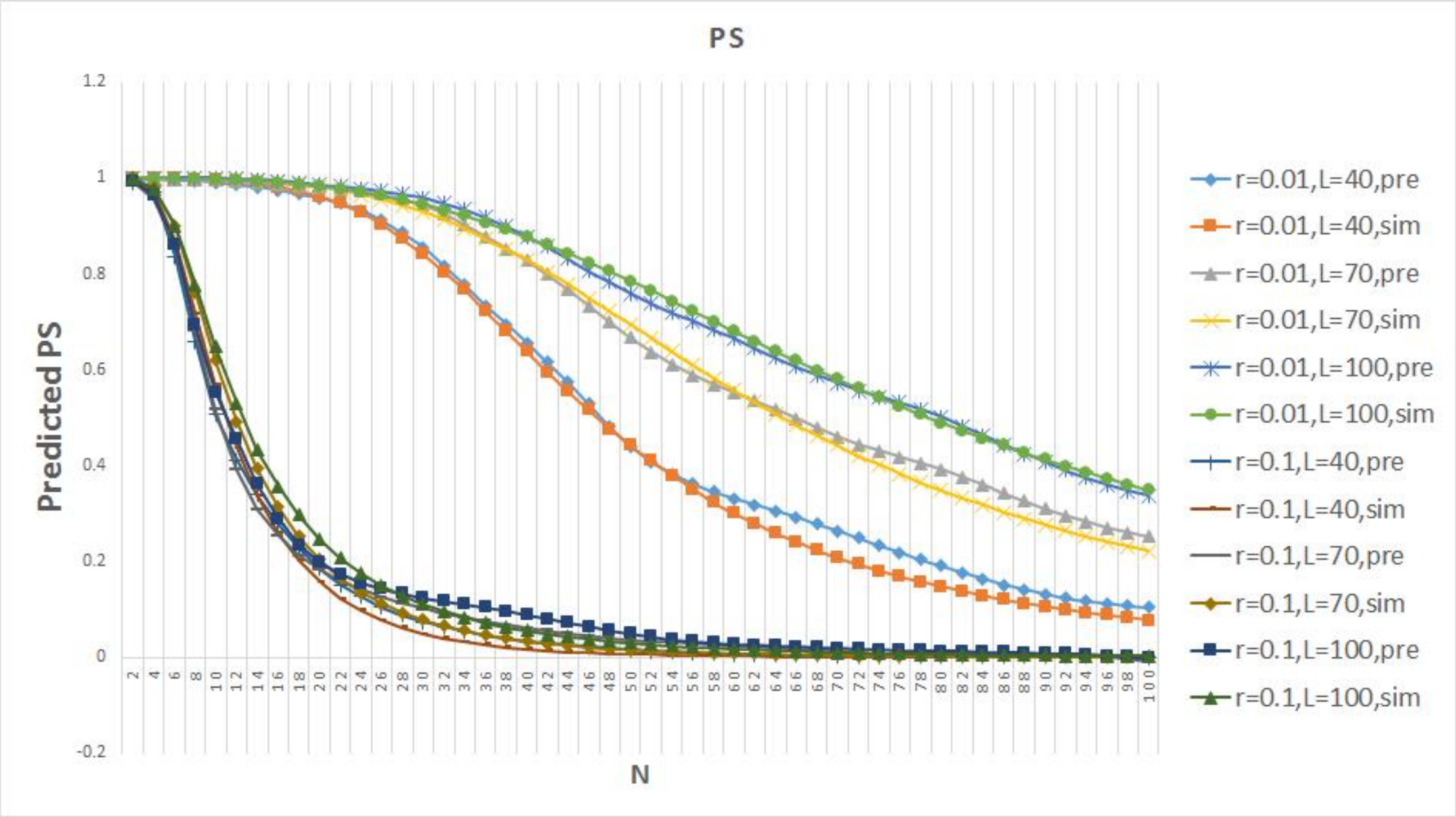

(b) Comparison of simulated and predicted *PS*

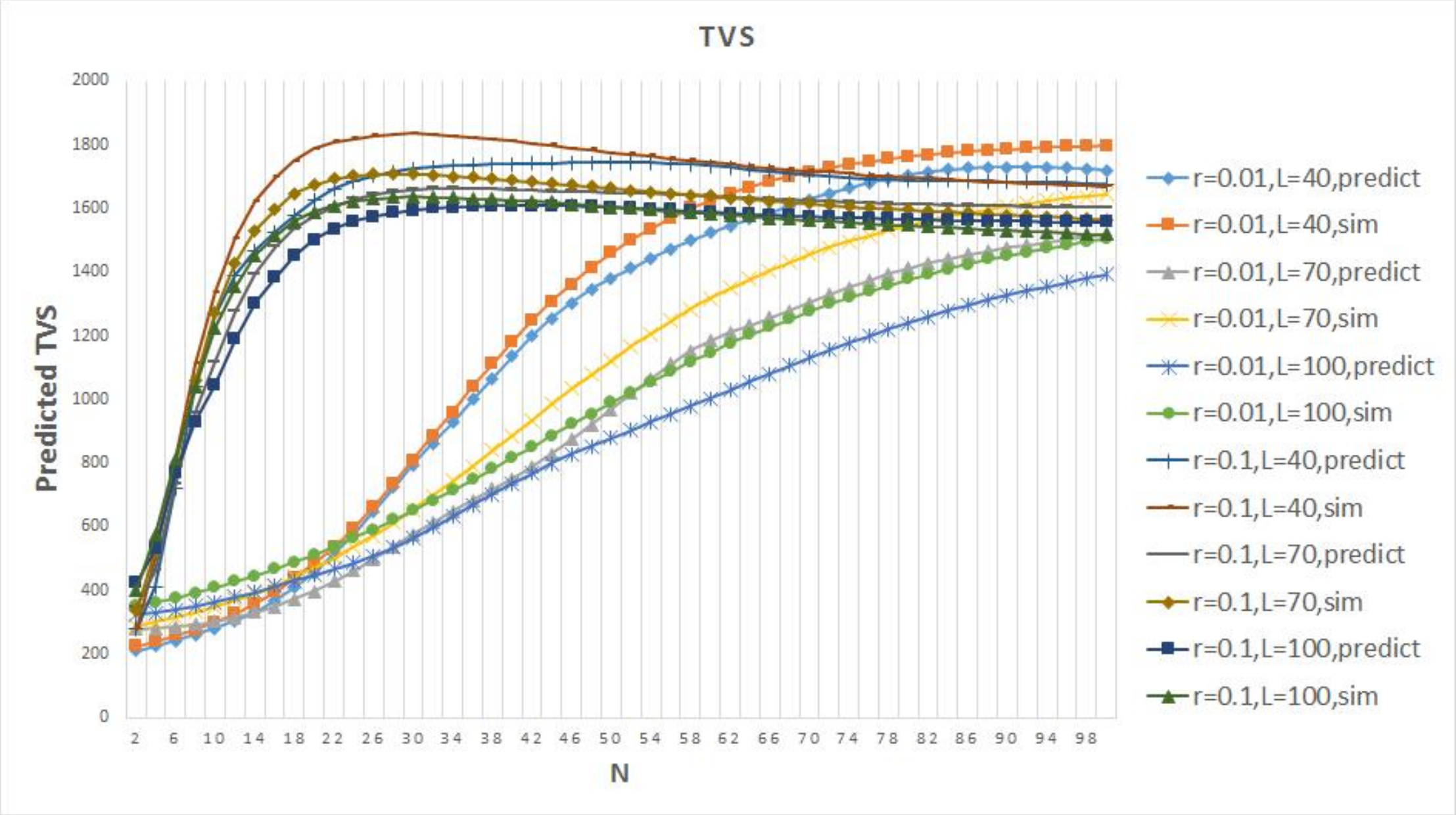

(c) Comparison of simulated and predicted *TVS*

**Fig. 14** Predict the average transmission rate using BP-ANNs

The prediction of *PS, N,* and *TVS* are partly close to the simulation results. Because the training set was derived from the analytical models, the gap between the analytical models and simulation affects the performance of the predicted results. Because the relationships of these parameters are interdependent, it is difficult to express one parameter by other parameters in a formula. Therefore, the BP-ANN model is a good choice for predicting some parameters of network using other specified parameters. This is helpful when users would like to balance the quality of service when deploying the sensor network.

# 5. Conclusion

Three Markov chain models were proposed to evaluate the performance of non-beacon IEEE 802.15.4 networks under different operation conditions, in which the ACK frame and retransmission limits are considered. The data packet arrival rate follows a Poisson process. The MAC buffer size, data frame length, node number, and packet arrival rate are the input parameters of the models. The expressions for eight parameters were derived from the analytical models. Based on the Monte Carlo simulation and analytical models, we compared the performance these eight parameters analytically and using simulation. The extensive simulation results show the accuracy of the analytical models with respect to throughput and reliability.

For unsaturated traffic, we found that these performance metrics are not only a function of the node number, data frame length, data frame arrival rate, MAC buffer size, CCA probability, and probability of sensing a busy channel, but are also interdependent with some performance parameters. The increase in MAC buffer size increases the average delay of packet transmission, and a large MAC buffer size does not improve the *PS* and *TH*. The increase of L may improve the *PS* and *TH* and reduce the average transmission delay. The increase of *N* and *r* raises the number of packets to be sent and quickly saturates the network. In unsaturated traffic, the non-beacon network performs well when N and r are small.

Furthermore, we used a BP-ANN to predict the performance of *N*, *PS*, and *TVS* by training on a dataset from the analytical results. The results of the BP-ANN models indicate that the R of the training set and testing sets are higher than 0.99. Indeed, the BP-ANN is a good tool for predicting the performance of a network given a quality of service requirement.

In our Markov chain models, the data packet arrival rate only follows the Poisson process. However, other distributions of data packet arrival rates such as uniform distribution should also be considered. The expressions of the parameters are still complex, and an approximation method is needed to simplify the expressions to reduce the computational complexity.

# References

[1] ZigBee Standards Organization: ZigBee Specification Document 053474r13. (2006). ZigBee Alliance.

[2] 6LoWPAN: The Wireless Embedded Internet, Shelby&Bormann. (2009). ISBN: 978-0-470-74799-5, (c) 2009 John Wiley & Sons Ltd.

[3] “Application Note: JN-AN-1035 Calculating 802.15.4 Data Rates” Jennic Ltd, South Yorkshire, UK. (2006). http://www.nxp.com/documents/application_note/JN-AN-1035.pdf. Accessed 25 January 2016.

[4] Burchfield, T. R., Venkatesan, S., & Weiner, D. (2007). Maximizing throughput in ZigBee wireless networks through analytical, simulations and implementations. *In Proc. Int. Workshop Localized Algor. Protocols WSNs*.

[5] Latre, B., Mil, P. D., Moerman, I., Dhoedt, B., Demeester, P., Dierdonck, N. V. (2006). Throughput and Delay Analytical of Unslotted IEEE 802. 15.4. *Journal of Networks*, 1(1), pp.

20--28.

[6] Lee, J. S. An experiment on performance study of IEEE 802.15. 4 wireless networks. Emerging Technologies and Factory Automation. (2005). *ETFA 2005. 10th IEEE Conference on. IEEE.*

[7] Zhu, J., Tao, Z., & Lv, C. (2011). Delay analytical for IEEE 802.15. 4 CSMA/CA scheme with heterogeneous buffered traffic. *Measuring Technology and Mechatronics Automation (ICMTMA), 2011 Third International Conference on. IEEE, 2011, 1: 835-845.*

[8] Anastasi, G., Conti, M., & Di Francesco, M. (2011). A comprehensive analytical of the mac unreliability problem in IEEE 802.15.4 wireless sensor networks. *IEEE Trans. on Industrial Informatics*, vol. 7, no. 1, pp. 52 - 65, Feb.

[9] Fernández-López, H., Macedo, P., Afonso, J. A., Correia, J. H., & Simões, R. (2010). Evaluation of the Impact of the Topology and Hidden Nodes in the Performance of a ZigBee Network. *In Sensor Systems and Software (pp. 256-271).* Springer Berlin Heidelberg.

[10] Sahoo, P. K., & Sheu, J. P. (2008). Modeling IEEE 802.15. 4 based wireless sensor network with packet retry limits. *In Proceedings of the 5th ACM symposium on Performance evaluation of wireless ad hoc, sensor, and ubiquitous networks (pp. 63-70). ACM.*

[11] Park, P., Di Marco, P., Soldati, P., Fischione, C., & Johansson, K. H. (2009). A generalized Markov chain model for effective analytical of slotted IEEE 802.15.4. *In Mobile Adhoc and Sensor Systems, 2009. MASS'09. IEEE 6th International Conference on (pp. 130-139). IEEE.*

[12] Jung, C. Y., Hwang, H. Y., Sung, D. K., & Hwang, G. U. (2009). Enhanced Markov chain model and throughput analytical of the slotted CSMA/CA for IEEE 802.15. 4 under unsaturated traffic conditions. *Vehicular Technology, IEEE Transactions on*, 58(1): 473-478.

[13] Shu, F., & Sakurai, T. (2011). A new analytical model for the IEEE 802.15. 4 CSMA-CA protocol. *Computer networks*, 55(11): 2576-2591.

[14] Wang, F., Li, D., & Zhao, Y. (2011). Analytical of CSMA/CA in IEEE 802.15. 4. *IET communications*, 5(15): 2187-2195.

[15] Al-Anbagi, I., Khanafer, M., & Mouftah, H. T. (2013). MAC finite buffer impact on the performance of cluster-tree based WSNs. *In Communications (ICC), 2013 IEEE International Conference on (pp. 1485-1490). IEEE.*

[16] Park, P., Di Marco, P., Fischione, C., & Johansson, K. H. (2013). Modeling and optimization

of the IEEE 802.15. 4 protocol for reliable and timely communications. *Parallel and Distributed Systems, IEEE Transactions on*, 24(3): 550-564.

[17] Yin, D., & Lee, T. T. (2013). Performance analysis of markov modulated IEEE 802.15. 4 beacon-enable mode. *Wireless networks*, 19(7): 1709-1724.

[18] Wijetunge, S., Gunawardana, U., & Liyanapathirana, R. (2014). Throughput analysis of IEEE 802.15. 4 beacon-enabled MAC protocol in the presence of hidden nodes. *Wireless Networks*, 20(7): 1889-1908.

[19] Kim, T. O., Park, J. S., Chong, H. J., Kim, K. J., & Choi, B. D. (2008). Performance analytical of IEEE 802.15. 4 non-beacon mode with the unslotted CSMA/CA. *IEEE Communications Letters*, 12(4): 238-240.

[20] Buratti, C., & Verdone, R. (2009). Performance analytical of IEEE 802.15. 4 non beacon-enabled mode. *Vehicular Technology, IEEE Transactions on*, 58(7): 3480-3493.

[21] Marco, P. D., Park, P., Fischione, C., & Johansson, K. H. (2010). Analytical modelling of IEEE 802.15. 4 for multi-hop networks with heterogeneous traffic and hidden terminals. In Global Telecommunications Conference (GLOBECOM 2010), 2010 IEEE (pp.1-6). IEEE.

[22] Srivastava, R., & Kumar, A. (2012). Performance analytical of beacon-less IEEE 802.15. 4 multi-hop networks. *In Communication Systems and Networks (COMSNETS), 2012 Fourth International Conference on (pp. 1-10). IEEE.*

[23] Feo, E., & Di Caro, G. A. (2011). An analytical model for IEEE 802.15.4 non-beacon enabled CSMA/CA in multihop wireless sensor networks. *Istituto Dalle Molle di Studi sull'Intelligenza Artificiale, Lugano, Switzerland, Tech. Rep*, 05-11.

[24] Wijetunge, S., Gunawardana, U., & Liyanapathirana, R. (2012). Throughput analytical of non-beacon enabled IEEE 802.15. 4 networks with unsaturated traffic. *In Communications and Information Technologies (ISCIT), 2012 International Symposium on (pp. 1177-1182). IEEE.*

[25] Samaras, I. K., & Hassapis, G. D. (2013). A Flexible Analytical Markov Model for the IEEE 802.15. 4 Unslotted Mechanism in Single-Hop Hierarchical Wireless Networks with Hidden Nodes. *Wireless personal communications*, 72(4): 2389-2424.

[26] Wang, F., Li, D., & Zhao, Y. (2009). Analysis and compare of slotted and unslotted CSMA in IEEE 802.15.4. *In Wireless Communications, Networking and Mobile Computing, 2009. WiCom'09. 5th International Conference on (pp. 1-5). IEEE.*

[27] Ling, X., Cheng, Y., Mark, J. W., & Shen, X. (2008). A renewal theory based analytical model for the contention access period of IEEE 802.15. 4 MAC. *Wireless Communications, IEEE Transactions on*, 7(6): 2340-2349.

[28] Lee, C. Y., Cho, H. I., Hwang, G. U., et al. (2011). Performance modeling and analytical of IEEE 802.15. 4 slotted CSMA/CA protocol with ACK mode. *AEU-International Journal of Electronics and Communications*, 65(2), 123-131.

[29] Shuaib, A. H., Mahmoodi, T., Aghvami, A. H. (2009). A timed Petri Net model for the IEEE 802.15. 4 CSMA-CA process. *In the Proceedings of IEEE 20th International Symposium on Personal, Indoor and Mobile Radio Communications*.

[30] Hwang, R. C., Hsu, P. T., Cheng, J., et al. (2011). The indoor positioning technique based on neural networks. *Signal Processing, IEEE International Conference on Communications and Computing (ICSPCC)*.

[31] Li, Y. Y., & Parker, L. E. (2008). Intruder detection using a wireless sensor network with an intelligent mobile robot response. In the Proceedings of IEEE Southeastcon.

[32] Nkwogu, D. N., Allen, A. R. (2012). Adaptive sampling for WSAN control applications using artificial neural networks. *Journal of Sensor and Actuator Networks*, 1(3), 299-320.

[33] Min, S., Xinyu, J., Jianting, L. (2010). An improved handover algorithm of Wireless Sensor Networks in high-speed mobile environment. *In the Proceedings of IEEE Information Science and Engineering (ICISE), 2nd International Conference*.

[34] Veena, K. N., Kumar, B. P. V. (2010). Convergecast in wireless sensor networks: a neural network approach. *In the Proceedings of IEEE 4th International Conference on Internet Multimedia Services Architecture and Application (IMSAA)*.

# Appendix A

In the CSMA/CA algorithm, the basic time unit is *aUnitBackoffPeriod* (= 20 symbols, where 1 symbol is equal to 16 μs). Here, we define the mini-slot to be 1 symbol, and a slot to be 20 symbols.

The sensor nodes execute one CCA before transmission. In Fig. A1, at the initial stage, two variables *NB* and *BE* are initialized. Variable *NB* is the number of backoff (periods) the CSMA-CA algorithm has been required to backoff while attempting the current transmission; this value is initialized to 0 before each new transmission attempt and its maximum value is *macMaxCSMABackoffs*, which is 4 by default. Variable *BE* is the backoff exponent, which is related to how many backoff periods a device will wait before attempting to access a channel. The initial value for *BE* is *macMinBE*, which is 3 by default, and the maximum value is aMaxBE, which is 5 by default. When the backoff periods end, the node starts a CCA, which costs 8 symbols. If the CCA fails because of busy channel sensing, the node will enter the next backoff stage. If *NB* > *macMaxCSMABackoffs*, the transmission fails. If the channel is idle when the nodes perform a CCA, the CCA is a success. The node state will turn from RX to TX in *aTurnaroundTime* (= 12 symbols).

In the ACK mode, the node waits to receive the ACK frame for $t_{ACK}$ after the data frame finishes sending, where $t_{ACK} \in$ [*aTurnaroundTime*, *aTurnaroundTime* + *aUnitBackoffPeriod*]. Here, we assume that $t_{ACK}$ is 20 symbols. The ACK frame is 11 bytes and the sending time is 22 symbols. If the node does not receive the ACK within *macAckWaitDuration* (= 54 symbols), the ACK frame is lost and the data frame transmission fails.

The node can retry to transmit the packet *aMaxFrameRetries* times (3 by default).

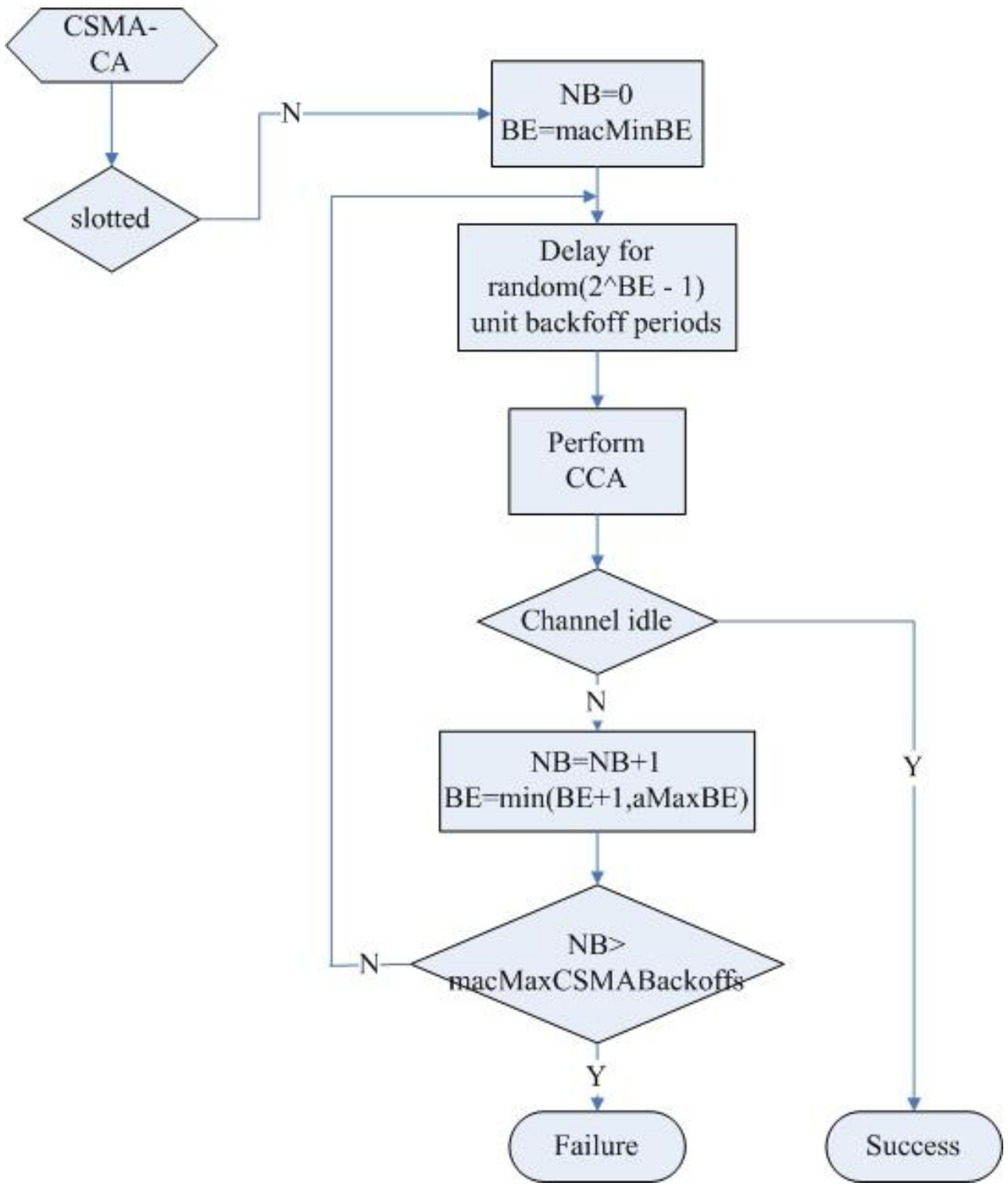

**Fig. A1** Unslotted CSMA-CA algorithm

For the unslotted CSMA/CA, we used the following default attributes.

*aMaxFrameRetries* 3

*macMinBE* 3

*macMaxBE* 5

*macMaxCSMABackoffs* 4

*macAckWaitDuration* 54 symbols

*aTurnaroundTime* 12 symbols

$t_{ACK}$ 20 symbols

# Appendix B

$$\pi(TA,m)=(1-a)\sum_{i=0}^{4}\pi(i,0,m),\ \ (m\in[0,3],i\in[0,4])\ \ (1)$$

$$\pi(TxC,m,1)=k\pi(TA,m),\ \ (m\in[0,3])\ \ (2)$$

$$\pi(TxC,m,n+1)=k\pi(TxC,m,n)\ \ (n\in[1,11])\ \ (3)$$

$$\pi(TxSuc,m)=k\pi(TxC,m,12),\ \ (m\in[0,3])\ \ (4)$$

$$\pi(Turn,m)=\pi(TxSuc,m),\ \ (m\in[0,3])\ \ (5)$$

$$\pi(AcW,m,1)=k\pi(Turn,m),\ \ (m\in[0,3])\ \ (6)$$

$$\pi(AcW,m,n+1)=k\pi(AcW,m,n),\ \ (n\in[1,11],m\in[0,3])\ \ (7)$$

$$\pi(AckSuc,m)=k\pi(AcW,m,12),\ \ (m\in[0,3])\ \ (8)$$

$$\pi(TF,m,0)=(1-k)\pi(TA,m),\ \ (m\in[0,3])\ \ (9)$$

$$\pi(TF,m,n+1)=(1-k)\pi(TxC,m,n+1)+\pi(TF,m,n),\ \ (n\in[0,11],m\in[0,3])\ \ (10)$$

$$\pi(turn,m)=\pi(TF,m,12),\ \ (m\in[0,3])\ \ (11)$$

$$\pi(AF,m,0)=\pi(turn,m)+(1-k)\pi(Turn,m),\ \ (m\in[0,3])\ \ (12)$$

$$\pi(AF,m,n+1)=\pi(AF,m,n)+(1-k)\pi(AcW,m,n+1),\ \ (n\in[0,11],m\in[0,3])\ \ (13)$$

# Appendix C

$$\pi(CW1)=y\cdot\theta\ \ (1)$$

$$\pi(CWi)=z\cdot\pi(CWi)=y\cdot z^{i-1}\cdot\theta\ \ \left(i\in[1,12]\right)\ \ (2)$$

$$\pi(TxSuc)=y\cdot z^{12}\cdot\theta\ \ (3)$$

$$\pi(IDLE3)=\pi(TxSuc)=y\cdot z^{12}\cdot\theta\ \ (4)$$

$$\pi(IWi)=y\cdot z^{12+i}\cdot\theta\ \ i\in[1,12]\ \ (5)$$

$$\pi(AckSuc) = z \cdot \pi(CW12) = y \cdot z^{25} \cdot \theta \quad (6)$$

$$\pi(TxFail) = (1-x-y)\pi(IDLE2) + (1-z)\sum_{i=1}^{12}\pi(CWi) = (1-x-y \cdot z^{12}) \cdot \theta \quad (7)$$

$$\pi(AckFail) = (1-z)[\pi(IDLE3) + \sum_{i=1}^{12}\pi(IWi)] = (1-z^{13}) \cdot y \cdot z^{12} \cdot \theta \quad (8)$$

$$\pi(wack) = \pi(TxFail) + \pi(AckFail) = (1-x-y \cdot z^{25}) \cdot \theta \quad (9)$$

$$\pi(IDLE1) = \pi(wack) + \pi(AckSuc) = (1-x) \cdot \theta \quad (10)$$